\documentclass[twocolumn,aps,floatfix,superscriptaddress,prl,showpacs]{revtex4}

\usepackage{graphicx}
\usepackage{amsmath,amssymb,amsfonts}
\usepackage{bm}
\usepackage{bibentry, natbib}
\usepackage[T1]{fontenc}
\usepackage[english]{babel}
\usepackage[latin1]{inputenc}
\usepackage{ifthen}
\usepackage{footnote}
\usepackage{subfigure}
\usepackage{setspace}

\usepackage{natbib}
\usepackage{latexsym,amsmath,amssymb,amsfonts,amsthm,enumerate,url}
\usepackage{hyperref}
\hypersetup{
  colorlinks=true,
  urlcolor=blue}

\usepackage{graphicx}
\usepackage{dcolumn}
\usepackage{bm}

\usepackage[usenames,dvipsnames]{color}

\usepackage{graphicx}
\usepackage{dcolumn}
\usepackage{bm}
\usepackage{latexsym,amsmath,amssymb,amsfonts,amsthm,enumerate}

\usepackage{hyperref}
\hypersetup{
colorlinks=true,
citecolor=blue,
linkcolor=red,
urlcolor=Green}

\renewcommand{\(}{\left (}
\renewcommand{\)}{\right )}
\renewcommand{\[}{\left [}
\renewcommand{\]}{\right ]}
\newcommand{\la}{\langle}
\newcommand{\ra}{\rangle}

\renewcommand{\t}{\theta }

\newcommand{\dg}{\dagger}

\renewcommand{\|}{\nonumber \\}

\renewcommand{\!}{\nonumber \\ & }
\newcommand{\eqa}[2][]{
\begin{align}
 #2 \label{#1}
\end{align}}

\newcommand{\be}{\begin{equation}}
\newcommand{\bea}{\begin{equation}\begin{array}l\displaystyle}
\newcommand{\ee}{\end{equation}}
\newcommand{\bes}{\begin{subequations}}
\newcommand{\esu}{\end{subequations}}

\renewcommand{\(}{\left (}
\renewcommand{\)}{\right )}
\renewcommand{\[}{\left [}
\renewcommand{\]}{\right ]}

\renewcommand{\t}{\theta }

\renewcommand{\|}{\nonumber \\}

\newcommand{\eq}[2][]{
\begin{equation}
#2 \label{#1} 
\end{equation}} 

\renewcommand{\!}{\nonumber \\ & }

\def\3pt#1#2#3{{\langle{#1}\vert{#2}\vert{#3}\rangle}}

\def\s {\sigma}
\def\be{\begin{equation}}
\def\ee{\end{equation}}
\def\EQ{\begin{equation}}
\def\EN{\end{equation}}
\def\bea{\begin{eqnarray}}
\def\eea{\end{eqnarray}}
\def\to{\rightarrow}

\def\sa{\hspace{0.1in}}

\def\sb{\hspace{0.2in}}
\def\sm{\hspace{0.15in}}

\begin{document}


\title{Boundary State in an Integrable Quantum Field Theory Out of Equilibrium }

\author{Spyros Sotiriadis}
\affiliation{Department of Physics, University of Pisa, Italy}
\affiliation{INFN, Pisa section, Italy}
\author{Gabor Takacs}
\affiliation{Department of Theoretical Physics, Budapest University of Technology and Economics, 
Hungary}
\affiliation{MTA-BME ``Momentum'' Statistical Field Theory Research Group, 
Hungary}
\author{Giuseppe Mussardo}
\affiliation{SISSA and INFN, Trieste, Italy}
\affiliation{The Abdus Salam ICTP, Trieste, Italy}


\begin{abstract}
We study a quantum quench of the mass and the interaction in the Sinh-Gordon model starting from a large initial mass and zero initial coupling. Our focus is on the determination of the expansion of the initial state in terms of post-quench excitations. We argue that the large energy profile of the involved excitations can be relevant for the late time behaviour of the system and common regularization schemes are unreliable. We therefore proceed in determining the initial state by first principles expanding it in a systematic and controllable fashion on the basis of the asymptotic states. Our results show that, for the special limit of pre-quench parameters we consider, it assumes a squeezed state form that has been shown to evolve so as to exhibit the equilibrium behaviour predicted by the Generalized Gibbs Ensemble.
\end{abstract}


\maketitle

{\em 1. Introduction}. Research in non-equilibrium processes of Quantum Field Theory (QFT) and their 
statistical mechanical properties constitutes a fast developing and widely applicable area of theoretical physics.  
The correct understanding of phenomena out of equilibrium not only plays a crucial role for our knowledge about as diverse topics as cosmology, heavy-ion collision experiments and cold atom systems (see, for instance, \cite{Rammer,Calzetta,Berges} and references therein), but it also poses purely theoretical questions in the subject of QFT itself. 

This is particularly true in the case of (1+1) dimensional integrable QFTs, i.e. systems which have an infinite number of local integrals of motion \cite{Mussardo}: in this case, connecting far-from-equilibrium dynamics at early times with the approach to equilibrium at late times may be a true challenge for the theory. In particular, the experimental evidence of lack of thermalization in a 1$d$ system of bosons with pointlike interactions \cite{Kinoshita} (a system described by a special limit of the integrable QFT of the Sinh-Gordon model \cite{KMTPRL,KMT}) led to the conjecture that quantum integrable systems exhibit equilibration to a \emph{Generalized Gibbs Ensemble} (GGE) rather than the usual Gibbs ensemble of thermal equilibrium \cite{Rigol}. 

The GGE is associated to a density matrix $$\rho_{GGE} \propto \exp(-\sum_i \lambda_i \mathcal{Q}_i)$$ that involves \emph{all} local integrals of motion $\mathcal{Q}_i$ of the integrable model, including the Hamiltonian. The validity of the GGE has been verified with a variety of different approaches and settings in many systems which can be mapped to free boson or fermion systems, even though such mappings are often highly non-trivial \cite{Cazalilla,Iucci,Iucci2,Essler,CEF1,CEF2,SE,CSC13a,CSC13b,Kormos13b,Andrei13c}. For genuine interacting integrable QFT there have been so far only a few studies \cite{Fioretto,FE13,CE,Pozsgay13a,Kormos13,RS13,Caux13,Pozsgay13b,Muss13,Andrei13a,Andrei13b}, and presently the most general result concerns the time-average of one-point functions of local operators, for which it was shown in \cite{Muss13} that their values can indeed be recovered by the GGE average.

In many non-equilibrium situations of a QFT, the future evolution of the system is entirely encoded into the specification of the initial state $|B \rangle$, also called {\em boundary state}. This is what happens, for instance, in the global \emph{Quantum Quench} (QQ) process, where a parameter of the Hamiltonian is abruptly changed at $t = t_0$ and the role of the boundary state is played by the ground state of the pre-quench Hamiltonian. 

The subject of this paper is the theoretical investigation of the boundary state closely related to Dirichlet boundary conditions of an interacting integrable QFT, in particular its determination according to basic principles.  An important issue of our analysis will concern  the proper treatment of the ultraviolet unbounded behaviour originally present in the expression of $| B \rangle $, a task that will lead us to a non-trivial set of equations involving the exact matrix elements of the field $\phi(x)$ on the asymptotic states.
 For convenience we will  present our main results through the simplest representative of these interacting theories, i.e. the Sinh-Gordon model based on the bosonic field $\phi(x)$, the generalisation to more complicated integrable QFT being straightforward. 

\begin{figure}[htbp]
\centering
\includegraphics[width= \columnwidth]{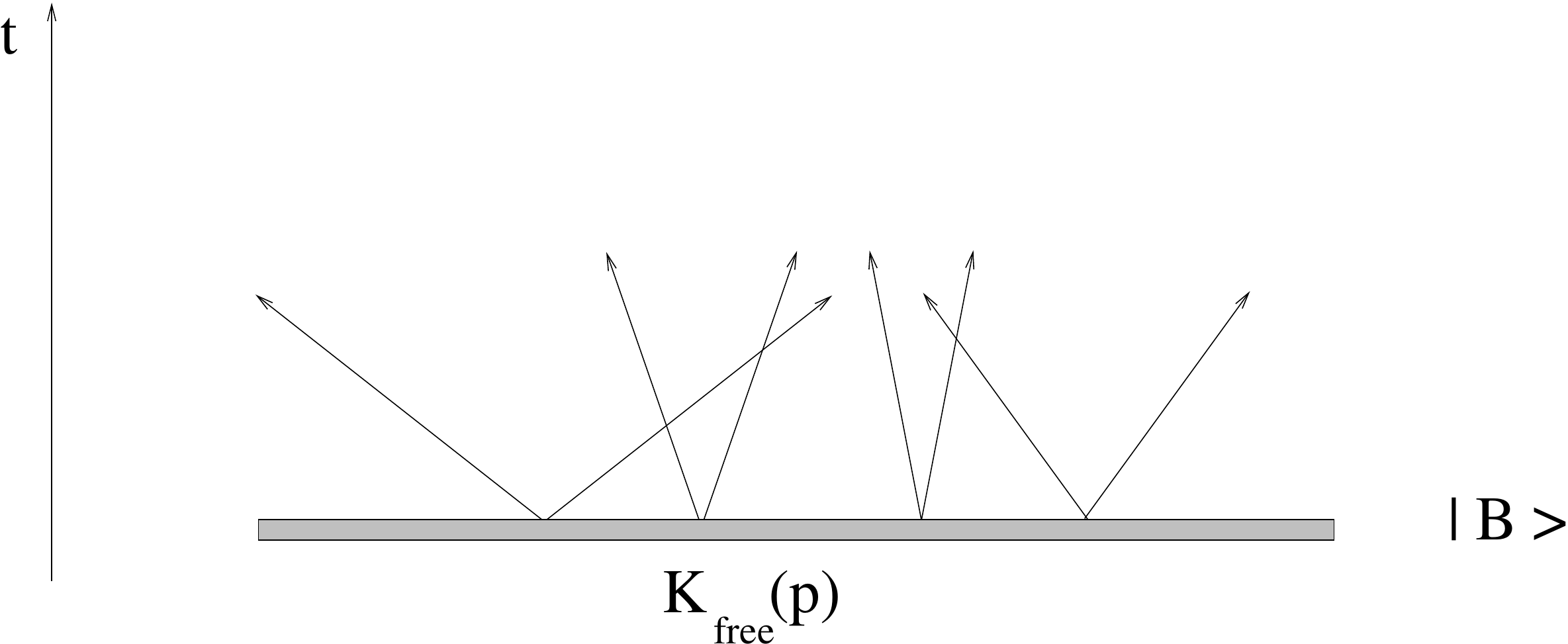}
\caption{ \label{Cooper} \small{With respect to the post-quench Hamiltonian, 
the boundary state $| B \rangle$ appears as a coherent superposition of an infinite number 
of pairs of particles with equal and opposite momentum.}}
\end{figure}

{\em 2. Boundary States}. In global QQ, $| B \rangle$ is the ground state of the pre-quench Hamiltonian and an important step for solving the subsequent out-of-equilibrium dynamics is to express this state in terms of the operators that create the particle excitations of the post-quench Hamiltonian. A familiar and simple example of this procedure, which will be important for our future considerations, is the quench process of the mass term $m_0 \rightarrow m$ of a free bosonic massive QFT \cite{CC1}: in this case, introducing  
$$
c_\pm(p) =\frac{1}{2} \left(\sqrt{\frac{E_0(p)}{E(p)}} \pm  \sqrt{\frac{E(p)}{E_0(p)}}\right) \,\,\,,
$$
with $E(p) = \sqrt{p^2+m^2}$ and $E_0(p) = \sqrt{p^2+m_0^2}$, and denoting by 
$(A_0(p), A^\dagger_0(p))$ and $(A(p), A^\dagger(p))$ the pre/post-quench annihilation and creation operators, these two sets 
are related by a Bogoliubov transformation 
\begin{eqnarray}
A_0(p) &=& c_+(p) A(p) + c_-(p) A^\dagger(-p) 
\label{bogoliubov}\\
A_0^\dagger(p) & = & c_+(p) A^\dagger(p) + c_-(p) A(-p) \nonumber 
\end{eqnarray} 
In this example, the boundary state is identified by the condition $A_0(p) | B \rangle =0$, which can be expressed in terms of 
the post-quench operators as 
\begin{equation}
\left[ c_+(p) A(p) + c_-(p) A^\dagger(-p)\right] | B \rangle = 0 \,\,\,. 
\end{equation}
The solution of this equation provides the sought expression of the boundary state in terms of the post-quench operators 
\eq[freeB]{| B_\text{free} \ra \sim \exp{\[-\int_0^\infty \frac{dp}{2\pi} \; K_\text{free}(p) \;  A^\dg(-p) A^\dg(p) \]} |\Omega\ra}   
where $| \Omega \rangle$ is the ground state of the post-quench Hamiltonian and 
\eq{K_\text{free}(p) = \frac{E_{0}(p)-E(p)}{E_{0}(p)+E(p)} \, .}
From the point of view of the post-quench system, the boundary state is then an infinite superposition of pairs of equal and opposite momentum, each of them weighted with the amplitude $K_\text{free}(p)$ 
(see Figure \ref{Cooper}). 

In the case of interacting integrable QFT (for simplicity we are considering integrable QFT with only one type of 
particle excitations), a generalization of this class of boundary states consisting of an infinite number of pairs of equal and opposite momentum is given by the general expression
\eq[sv]{ | B \ra \sim \exp\(\int_0^\infty \frac{d\t}{2\pi} \; K(\t) Z^\dg(-\t) Z^\dg(\t) \) |\Omega\ra }
where $Z^\dg(\t)$ are creation operators of the post-quench Hamiltonian, $|\Omega\ra$ is its ground state and the variable $\t$ conveniently parameterizes the dispersion relation of the particle excitations of mass $m$, given by $E= m \cosh\t$ and $p=m \sinh\t$.  
The operators $Z(\t)$ and $Z^\dg(\t)$ provide a complete basis of the Hilbert space of the post-quench integrable QFT and satisfy the Zamolodchikov-Faddeev algebra
\begin{align}
& Z(\t_1) Z(\t_2)  = S(\t_1-\t_2) Z(\t_2) Z(\t_1) \label{ZFa} \\
& Z(\t_1) Z^\dg(\t_2)  = S(\t_2-\t_1) Z^\dg(\t_2) Z(\t_1) + 2 \pi \delta(\t_1-\t_2) \nonumber
\end{align}
that involves the exact two-body $S$-matrix $S(\t_1-\t_2)$, function of the rapidity differences. Boundary states of the exponential form (\ref{sv}) have been considered in \cite{Fioretto,Pozsgay}, where it was shown that they automatically lead to equilibration of one-point observables according to the GGE (in agreement with the general analysis done in \cite{Muss13}). Such a class of exponential states includes the important examples of \emph{integrable boundary states}, i.e. boundary states that respect the integrability of the bulk \cite{Ghoshal}, like the Dirichlet states that can be prepared by a special QQ where the mass parameter $m_0$ of the pre-quench Hamiltonian is sent to infinity. 

While it is presently not known whether a general QQ in an integrable QFT leads to an exponential state (\ref{sv}), we will argue that this may be the case for a natural class of quench processes in the Sinh-Gordon model. Before facing this aspect of the problem, let us  discuss another important issue related to boundary states.  
 
{\em 3. The problem of ultraviolet behaviour}. The Dirichlet state $|D\rangle$, like all other known integrable boundary states, suffers from 
an ultraviolet unbounded behaviour. For example, taking literally the expression that comes out in the limit $m_0 \rightarrow \infty$ of the free bosonic case (\ref{freeB}), one realizes that the corresponding amplitude $K_D(p)$, associated to an idealized Dirichlet state, is a constant and does not decay sufficiently fast for large momenta. This is easily understood since excitations on the initial state are cut-off by $m_0$ which in this case is taken to be infinite. Similar unbounded behaviour is present in any other integrable QFT and, as a result, gives rise to divergent expectation values. 

When the post-quench Hamiltonian is a Conformal Field Theory, a cure of this
problem was proposed by Cardy and Calabrese \cite{CC1,CC2}, who assumed a large but not infinite $m_0$ and made use of the concept of `extrapolation length' $\tau_0$, already known from boundary Renormalization Group (RG) theory, to account for the difference of the actual initial state from the idealized Dirichlet state. The parameter $\tau_0$ is a `small' regularization parameter that plays the role of the inverse of an exponential cut-off  and depends on the initial parameters: chosen to be of the order $1/m_0$, it turns out to give indeed a good approximation of the initial state when the post-quench system is critical. 

In the case of massive post-quench theories, the obvious generalisation of the above idea is to replace $ K_D(\t) $ by $ K_D(\t) e^{-2 E(\t) \tau_0} $ and to postulate that $\tau_0$ is still of order $1/m_0$ as in the conformal case. In order to check the validity of this assumption and estimate a suitable value for $\tau_0$, one may choose an observable and equate its expectation values in the exact and in the approximate initial state. If the $\tau_0$ regularization were consistent, this estimate should be independent of the choice of observable under consideration. However this turns out not to be true, as shown in detail in the Supplementary Material.  For instance, for a mass quench in free bosonic theory, choosing as observable the operator $\phi^2(x)$, one arrives at the scaling relation
\eq[texttau_0]{ \tau_0 = \frac{\pi}{2e^{\gamma}} m_0^{-1} \approx 0.881938 \; m_0^{-1} \,\,\,,}
where $\gamma$ is the Euler-Mascheroni constant. Although this goes as in the conformal case,  $\tau_0 \sim 1/m_0$
\cite{CC1,CC2}, the prefactor is however different. A similar scaling law, but with another prefactor, is obtained choosing as observable the Hamiltonian: in this case one arrives at 
\eq[texttau_0b]{\tau_0 = \frac{\pi}{2\sqrt{3}} m_0^{-1} \approx 0.906900 \; m_0^{-1} \,\,\,.} 
Even though the two scaling laws (\ref{texttau_0}) and (\ref{texttau_0b}) are very close numerically, the impossibility to arrive to a universal expression of $\tau_0$ is nevertheless a flaw of the present regularization scheme. This discrepancy can be interpreted as an indication that the effect of higher energy excitations present in the initial state cannot be incorporated in an appropriate and unique definition of an energy cut-off: observables that weigh differently the effect of low and high energy excitations can then reveal different ultraviolet behaviour. 

A way out of these difficulties is to assume $\tau_0$ to be not a constant but a quantity that depends on $p$ itself.  In fact, such a dependence is perfectly justified from the point of view of boundary RG, according to which the actual boundary state may be constructed involving any boundary irrelevant operator, as recently discussed in \cite{Cardy12a,Cardy12b}. 
In this approach, the introduction of the extrapolation length $\tau_0$ amounts to a perturbation of the boundary state generated by the bulk Hamiltonian i.e. the state becomes $e^{-H\tau_0}|D\rangle$. In addition to the latter, one must in general introduce a different $\tau_0$ for each bulk conserved charge $\mathcal{Q}_s$, which are indeed boundary irrelevant operators. This would lead to a regularised initial state obtained by  $e^{-\sum_s \mathcal{Q}_s \tau_{0,s}}|D\ra$, which turns out to be still of the form (\ref{sv}) but with a $\tau_0$ that is a momentum-dependent function. This is because all charges $\mathcal{Q}_s$ of an IFT can be put in the form $\int d\theta e^{s\theta} Z^\dagger(\theta) Z(\theta)$ \cite{Mussardo}. However the problem of how to determine the suitable function $\tau_0(\theta)$ or equivalently $K(\theta)$ remains.

In the following we will study a QQ in the Sinh-Gordon (shG) model in which we start from a large but not infinite mass $m_0$ and use the exponential form (\ref{sv}) as an Ansatz, providing a series of arguments for such a choice. We then derive, from first principles, a sequence of integral equations that must be satisfied by the function $K(\theta)$ and propose a solution based on an analytical approximation which we verify numerically with a high level of accuracy. This provides a posteriori a non-trivial check of the validity of our initial Ansatz (\ref{sv}).

{\em 4. The Sinh-Gordon Model.} 
The shG Hamiltonian is 
\eq{H = \frac12 \pi^2(x) + \frac12\left(\frac{\partial\phi}{\partial x}\right)^2 + \frac{\mu^2}{g^2}\left(\cosh g\phi-1\right)}
where $\phi=\phi(x,t)$ is a real scalar field, $\pi(x)$ its conjugate momentum, $\mu$ the mass and $g$ the coupling constant. In this integrable field theory there is only one type of particle with physical renormalized mass $m$ given by $m^2 = \mu^2 {\sin \alpha \pi}/{\alpha \pi} $ 
where $\alpha$ is the dimensionless renormalized coupling constant $\alpha = { g^2}/{(8\pi+ g^2)}$. 
Particle scattering is fully determined by the two-particle $S$-matrix given by \cite{Fring:1992pt,KM} 
\eq[eq_sinh_S]{ S(\t)=\frac{\sinh\t -i\,\sin \alpha\pi}{\sinh\t +i\,\sin \alpha\pi} \,\,\,,}
where $\t $ is the rapidity difference between the particles.

Let us consider a QQ in the shG model starting from a large initial mass $m_0$ and, for reasons that become clear soon, zero interaction $\alpha_0 = 0$:  the final quantities are finite values of $m$ and $\alpha$. Such a quench may be regarded as made of a sequence of processes: an initial quench of the mass in free bosonic theory, $m_0 \rightarrow m$, swiftly followed by a switching on of the coupling, $\alpha_0 \rightarrow \alpha$. 

To determine the boundary state $| B \ra$ for this QQ in terms of the post-quench Hamiltonian, let us  use the condition that $| B \rangle$ is annihilated by the pre-quench annihilation operator $Z_0(p)$ \cite{SFM}. The choice of the initial coupling value $\alpha_0=0$ is particularly convenient because in this case $Z_0(p)$ is just the annihilation operator of the free bosonic theory, easily expressible in terms of the physical field operator $\phi(x)$ and its conjugate momentum $\pi(x)$ as 
\eq{Z_{0}(p) = \sqrt{\frac{E_{0p}}{2}} \; \( \tilde\phi(p) + i \frac{\tilde\pi(p)}{E_{0p}} \) \,\,\,, }
where $\tilde{\phi}(p) \equiv \int dx \; e^{-ipx} \phi(x)$ is the Fourier transform of $\phi(x)$ and $\tilde{\pi}(p)$ of $\pi(x)$. Since 
in a QFT we have $\pi = \dot\phi = -i[\phi,H]$, we arrive to the following equation
\eq[3]{ ( \tilde\phi(p) + {[\tilde\phi(p),H]}/{E_{0p}} ) \, |B \ra = 0 \,\,\,.}
To make progress in the solution of this equation, 
let us  first expand the state in the post-quench basis in the most general way
\eqa[psi1]{ |B \ra = & \Big ( 1+\sum_{s=2 \atop s \text{ even}}^\infty \prod_{r=1}^s  \int\limits_{-\infty}^{+\infty} \frac{d\t_r}{2\pi} \; 2\pi \delta{\textstyle\({ \sum_{r=1}^s p(\t_r) }\)} \; \| 
& \tilde K_s( \{\t_r\} ) \Big ) \; |\t_1,...,\t_s\ra }
where $|\t_1,...,\t_s\ra  \equiv Z^\dg (\t_1)... Z^\dg (\t_s)|\Omega\ra $ ( $\t_1 \geq \t_2 \geq \t_3 \ldots $) is the post-quench eigenstate containing $s$ particles with rapidities $\t_1,...,\t_s$ and $p(\t_r)=m\sinh\t_r$ is the momentum corresponding to rapidity $\t_r$. 

Additional constraints on $| B \rangle$ come by exploiting the symmetries of the quench process and the boundary state. 
Since this is the ground state of the pre-quench free Hamiltonian, it is invariant under parity and translation invariance. Moreover both symmetries are preserved by the quench process: hence, for parity reason, the sum runs over even integer numbers of particles only,  while, for translation invariance, each term in the sum has zero total momentum, as ensured by the $\delta$-function.

Applying suitable test states on the left of (\ref{3}), we can derive integral equations satisfied by the amplitudes $\tilde K_s$ of the excitations present in $| B \ra$. However our investigation drastically simplifies if we assume that the state is of the exponential form (\ref{sv}). If we apply first the assumption that the state consists only of pairs of particles with opposite rapidities, we have
\eqa[psi2]{ |B \ra = \sum_{s=0}^\infty \prod_{r=1}^s  \int\limits_{-\infty}^{+\infty} \frac{d\t_r}{2\pi} \, K_{s}( \t_1,...,\t_s ) |{-\t_1},\t_1,...,{-\t_s},\t_s\ra }
where, due to the algebra (\ref{ZFa}), the amplitudes $K_{s}$ satisfy the properties 
\eqa{& K_{s}(...,-\t_i,...) = S(-2\t_i) K_{s}(...,\t_i,...) \nonumber \,\,\,,\\
& K_{s}(...,\t_i,...,\t_j,...) = K_{s}(...,\t_j,...,\t_i,...)\nonumber \,\,\,.} 
Assuming further that the state is of the more special form (\ref{sv}), the amplitudes $K_{s}$ are related to each other by
\eq{ K_{s}(\t_1,...,\t_s ) = \frac{1}{s!} \, \prod_{r=1}^s K_1(\t_r)\,. } 
The plausibility of such an Ansatz comes from a series of reasons: first of all, from the vanishing of the expectation values on the state $| B \rangle$ of all infinite conserved charges ${\mathcal Q}_a^-$ ($a=1,3,5,\ldots $) of the Sinh-Gordon model which are odd under parity transformation. Indeed, if $P$ is the parity operator which is conserved in the quench process, then $P Q^-_a P = -Q^-_a$ and since $|B\rangle$ is an even state, $P |B \rangle = |B\rangle$. Therefore 
$$ \la B | {\mathcal Q}_a^- | B \rangle = 
\la B | P^2 {\mathcal Q}_a^- P^2 | B\rangle = - \la B | {\mathcal Q}_a^- | B \rangle = 0\,.$$

Since on the asymptotic states such charges act as ${\mathcal Q}_a^- |\t_1,...,\t_n\ra = \sum_{k=1}^n \sinh(a\t_k) \, |\t_1,...,\t_n\ra 
$ a pair-wise structure of the boundary state automatically guarantees the vanishing of the expectation values on the state $| B \rangle$.  
Secondly, imagine to realize the overall quench in terms of a sequence of quenches, the first QQ$_1$ in which we change only the mass (at $\alpha_0=0$), the second QQ$_2$ in which we switch on the coupling. After QQ$_1$, the resulting boundary state is $|B_\text{free} \rangle$ given in (\ref{freeB}), which is made of pairs of equal and opposite particles created by the free operators $Z_0^\dagger(p)$ with mass $m$. After QQ$_2$, where we have switched on the coupling constant $\alpha$, the infinite number of pairs present in $|B_\text{free} \rangle$ start interacting between them. However the interaction provided by the Sinh-Gordon model cannot create or destroy particles since it is integrable and when the particles cross each other, they just experience a time-delay dictated by the elastic $S$-matrix given in (\ref{eq_sinh_S}). 
It is therefore conceivable that the only effect of interaction is to "dress" both the free particle amplitude 
$K_\text{free}(\theta) \rightarrow K(\theta)$ and the free creation operators $Z_0^\dagger(p) \rightarrow Z^\dagger(p)$, 
preserving though the pair-wise structure of the boundary state.    

Assuming the validity of the pair-wise structure of the initial state and the exponentiation of the amplitudes, i.e. assuming the form (\ref{sv}), let us  start our analysis from the limit $m_0\to\infty$ which corresponds to the Dirichlet state $|D\ra$ satisfying the condition 
\eq[3a]{\tilde\phi(p) |D\ra = 0 \,\,\,.}
Such a boundary state is known to be of the exponential form (\ref{sv}) with amplitude $K_D(\t)$ given by \cite{Ghoshal2}
\eq[KD]{K_D(\t) = i \tanh{(\t/2)}\(\frac{1+\cot (\pi \alpha/4 - i \t  /2)}{1-\tan (\pi \alpha/4 + i \t  /2)}\)\,\,\,.}
Such a known case provides a non-trivial check of the approach we are going to propose. Indeed, if we now take as test state an arbitrary 1-particle excitation $\langle\t |$ applied to the left of (\ref{3a}), substitute (\ref{sv}) and expand, we obtain in this way the following integral equation that must be satisfied by the amplitude $K_D(\t)$
\eqa[KDseries]{ & \sum_{s=0}^\infty \frac{1}{s!} \(\prod_{i=0}^s \int_{C_i}  \frac{d\t '_i}{2\pi} K_D(\t '_i)\) \!
\times F_{2s+1}(\t + i\pi,-\t'_1,\t'_1,...,-\t'_s,\t'_s) = 0\,\,\,,}
where $F_n(\{\t _j\})$ are the matrix elements (the so-called \emph{Form Factors}) of the field $\phi$ defined by $ F_n(\{\t _j\}) \equiv \la 0 |\phi| \{\t _j\} \ra $.  In the derivation of (\ref{KDseries}) we have exploited both the crossing symmetry and the analytical properties of QFT \cite{Smirnov,Mussardo} which have allowed us to write the matrix elements $\la \t | \phi | \{-\t'_i,+\t'_i\} \ra$ in terms of the Form Factors above. The exact expressions of the Form Factors of the Sinh-Gordon model were computed in \cite{Fring:1992pt,KM} (for convenience, their exact expressions can be found in the Supplementary Material). Moreover since the numerical value of the $F_n$'s  decreases with the order $n$, the series (\ref{KDseries}) shows a fast convergent behaviour and can be approximated to the desired order of accuracy simply restricting to the lowest terms. 

\begin{figure}[htbp]
\centering
\includegraphics[width= \columnwidth]{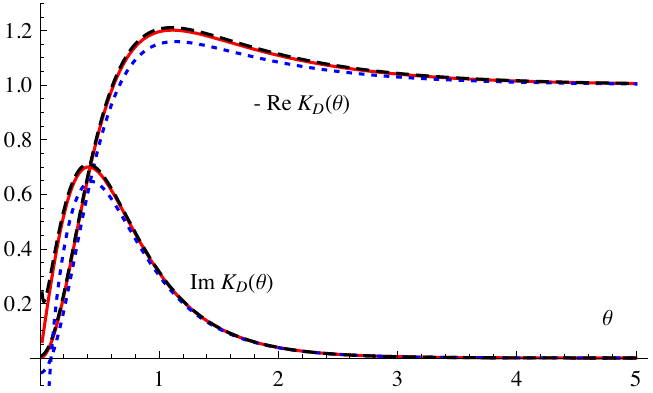}
\caption{ \label{figKD} \small{Plot of the numerical solution of (\ref{IE1}) truncated after the first 3 (blue dotted line) or 5 terms (black dashed line), along with the analytical result (\ref{KD}) (red full line) for $m=1, \alpha=0.4$.}}
\end{figure}

There is however a technical issue to take care of: since the Form Factors have poles whenever an in- and an out-rapidity coincide, one needs to choose a suitable prescription on how to pass around the poles, in order that the equation above makes sense. This prescription is encoded in the integration contours $C_i$ which can be determined by means of a \emph{finite volume regularization} \cite{PozsgayTakacs,KormosPozsgay,Pozsgay:2007kn,Pozsgay:2007gx} (details are discussed in the Supplementary Material). Once this prescription is implemented, the first few terms of the series give as a result the equation
\begin{widetext}
\eqa[IE1]{ & 0 = F_{1} + \frac{1}{2}F_{1}K_D(\theta)\( 1+S(-2\theta)\) + \frac{1}{2}\int\limits_{-\infty+i\epsilon}^{+\infty+i\epsilon}\frac{d\theta'}{2\pi}F_{3}(\theta+i\pi,-\theta',\theta')K_D(\theta')  \|
 & + \frac{1}{4}\int\limits_{-\infty}^{+\infty}\frac{d\theta'}{2\pi} \Big(S(-2\t)K_D(\t)+S(2\t)S(\t-\t')S(\t+\t')K_D(-\t)\Big) 
 F_{3}(-\theta,-\theta',\theta')K_D(\theta')  \|
 & + \frac{1}{8}\int\limits_{-\infty+i\epsilon}^{+\infty+i\epsilon}\frac{d\theta'_1}{2\pi}\int\limits_{-\infty+i\epsilon}^{+\infty+i\epsilon}\frac{d\theta'_2}{2\pi}F_{5}(\theta+i\pi,-\theta'_1,\theta'_1,-\theta'_2,\theta'_2) K_D(\theta'_1) K_D(\theta'_2)  + \dots}
\end{widetext}
In Fig.~\ref{figKD} we plot the numerical solution of (\ref{IE1}) when we keep its first three and five terms, along with the analytical result (\ref{KD}). As shown in the Figure, the agreement is quite satisfactory even when the series is truncated up to the first three terms and it improves significantly once the next two terms are included.
Using multi-particle test states applied on the left of (\ref{3a}), one obtains a series of equations that must all be satisfied by the amplitude $K_D(\t)$. More details of such computations will be presented elsewhere \cite{long_version}. 

Supported by this positive check for the case $m_0 \rightarrow \infty$, let us now address the problem of determining the 
amplitude $K(\theta)$ in the case of large but finite $m_0$. The equation that defines the initial state is now (\ref{3}). Considering a 1-particle test state as before and substituting (\ref{sv}), we find, after some algebra, that the new equation is
\eqa[Kseries]{ & \sum_{s=0}^\infty \frac{1}{s!} \(\prod_{i=0}^s \int_{C_i}  \frac{d\t '_i}{2\pi} K(\t '_i)\) \, \( \textstyle{ E_0(\t) - E(\t)  + \sum_{i=1}^s E(\t'_i) } \) \!
\times F_{2s+1}(\t + i\pi,-\t'_1,\t'_1,...,-\t'_s,\t'_s) = 0}
which, after a truncation of the series to the same order as before, becomes

\

\begin{widetext}
\eqa[K]{ 0 = & F_{1} \( \frac{E_{0}(\t)-E(\t)}{E_{0}(\theta)+E(\theta)} \) + \frac{1}{2}F_{1}K(\theta)\( 1+S(-2\theta)\)  + \frac{1}{2}\int\limits_{-\infty+i\epsilon}^{+\infty+i\epsilon} \frac{d\theta'}{2\pi} \(\frac{E_{0}(\theta)-E(\theta)+2E(\theta')}{E_{0}(\theta)+E(\theta)}\)  F_{3}(\theta+i\pi,-\theta',\theta')K(\theta')  \|
 & + \frac{1}{4}\int\limits_{-\infty}^{+\infty}\frac{d\theta'}{2\pi} \(\frac{E_{0}(\theta)+E(\theta)+2E(\theta')}{E_{0}(\theta)+E(\theta)}\) F_{3}(-\theta,-\theta',\theta')
  \Big(S(-2\t)K(\t)+S(2\t)S(\t-\t')S(\t+\t')K(-\t) \Big) K(\t') \|
 & + \frac{1}{8}\int \limits_{-\infty+i\epsilon}^{+\infty+i\epsilon}\frac{d\theta'_1}{2\pi}\int\limits_{-\infty+i\epsilon}^{+\infty+i\epsilon}\frac{d\theta'_2}{2\pi} \(\frac{E_{0}(\theta)-E(\theta)+2E(\theta'_1)+2E(\theta'_2)}{E_{0}(\theta)+E(\theta)}\) F_{5}(\theta+i\pi,-\theta'_1,\theta'_1,-\theta'_2,\theta'_2)  \, K(\theta'_1)K(\theta'_2)  + \dots}
\end{widetext}

One way to obtain the solution of this equation is to notice that, for a smooth function $K$, the integral of the first line is dominated by the contribution of the kinematical poles of the Form Factor at $\t=\pm \t' $. At these poles, the prefactor $({E_{0}(\t )-E(\t )+2E(\t'))}/{(E_{0}(\t)+E(\t))}$ 
of the integration kernel becomes equal to unit. This suggests the approximate solution
\eq[ans]{ K(\t) \approx K_D(\t)  \(\frac{E_{0}(\theta)-E(\theta)}{E_{0}(\theta)+E(\theta)}\) }
since then the first line of (\ref{K}) becomes approximately the same as the first line of (\ref{IE1}). The second and third lines contribute only small corrections to the solution. 

\begin{figure}[htbp]
\centering
\includegraphics[width= \columnwidth]{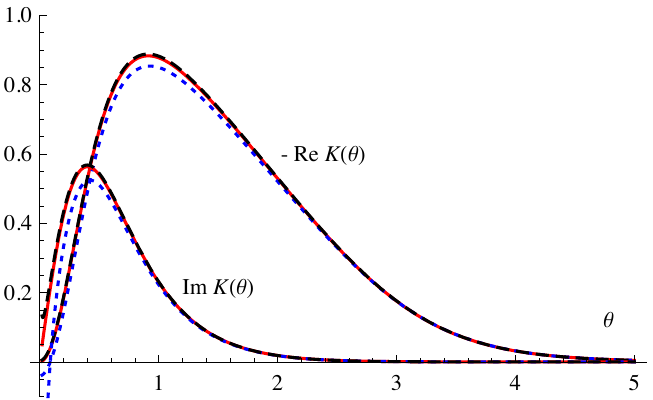}
\caption{ \label{figM} \small{Plot of the numerical solution of (\ref{K}) truncated after the first 3 (blue dotted line) or 5 terms (black dashed line), along with the analytical Ansatz (\ref{ans}) (red full line) for $m=1, m_0=10, \alpha=0.4$. }} 
\end{figure}

The correctness of our approximate solution can be verified numerically. Fig.~\ref{figM} shows a typical plot of a numerical solution of (\ref{K}) truncated to the 3rd or 5th terms, along with the Ansatz (\ref{ans}), for some values of the ratio $m_0/m$ and interaction $\alpha$. The agreement is quite satisfactory, even when we include the contribution of the second and third lines. Further comparative plots for a wide range of parameter values will be presented elsewhere \cite{long_version}. 

Obviously, the proposed solution (\ref{ans}) is expected to be more accurate when the parameters $m_0,m$ and $\alpha$ are such that the domination of the poles is more prominent and the higher order terms of the series give smaller contributions. The second condition is satisfied for example when both masses $m_0$ and $m$ are large or when the interaction $\alpha$ is small.

In analogy to the Dirichlet state case, using multi-particle test states we can derive a series of equations that must be satisfied by the amplitudes $K_s$ of the initial state in the form (\ref{psi2}). Based on the same combination of arguments used above (truncation of the form factor series and pole dominance of the integrals), it is possible to show that also these equations reduce approximately to the ones corresponding to the Dirichlet case when the $K_s$ are chosen to be $K_{s}(\t_1,...,\t_s ) = 1/s! \, \prod_{r=1}^s K_1(\t_r)$ with $K_1(\t) \approx K_D(\t) K_\text{free}(\t)$. In this way the exponential form of the Dirichlet state leads also to approximate exponentiation of the QQ initial state.

\begin{figure}[htbp]
\centering
\includegraphics[width= \columnwidth]{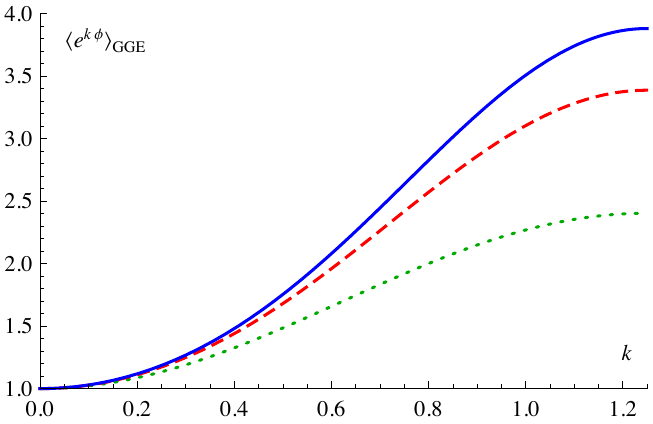}
\caption{ \label{figGGE} \small{Plot of the GGE prediction $\la e^{k\phi}\ra_{\text{GGE}}$ as a function of $k$ ($m=1, m_0=10, \alpha = 0.4$). The three curves represent partial sums of the series (\ref{gge_pred}): the dotted (green) line corresponds to the sum of the first two terms, the dashed (red) to the first three and the solid (blue) to the first four terms.}}
\end{figure}

\emph{5. Observables at large times.}  
According to the analysis done above, a first order approximation of the initial state $|B\ra$ for the QQ under consideration is given by the exponential form (\ref{sv}) with amplitude $K(\theta)$ given by (\ref{ans}). This decays for large momenta as a power law ($\sim p^{-2}$) and ensures a smooth ultraviolet behaviour through a momentum dependent $\tau_0$-regularization. An initial state of this form belongs to the class studied in \cite{Muss13,Fioretto,Pozsgay} and therefore at least the one-point functions of local observables equilibrate according to the GGE: their long time values are given by
\eqa[gge_pred]{ \mathcal{O}(x,t\to\infty) & = \sum_{n=0}^\infty \frac1{n!} \int\limits_{-\infty}^{+\infty} \prod_{i=1}^n \frac{d\t_i}{2\pi} \( \frac{|\bar{K}(\t_i)|^2}{1+|\bar{K}(\t_i)|^2} \) \times \|
& \la \t_n,...,\t_1 | \mathcal{O}(x) |\t_1,...,\t_n \ra_c }
where $\bar{K}$ is given by the solution of the generalised Thermodynamic Bethe Ansatz (TBA) equation
\eq{|\bar{K}(\t)|^2 = |K(\t)|^2 \exp\[ \int \frac{d\t'}{2\pi} \, \varphi(\t-\t') \log \(1+|\bar{K}(\t')|^2 \)\]}
where $\varphi(\t)\equiv -i \, d(\log{S(\t)})/d\t$, as explained in \cite{Fioretto}. From the above equations we can calculate numerically, for instance, the GGE prediction for the operator $\exp (k\phi)$ from which one can also derive all field fluctuations $\phi^{2n}$ by differentiation with respect to $k$ at $k=0$. In Fig.~\ref{figGGE}, the three curves represent three successive partial sums of the series (\ref{gge_pred}). The convergence of this series is particularly fast near the point $k=0$, even though to compute the higher moments $\phi^{2n}$ with sufficient accuracy one needs to employ more terms of the series.

\emph{6. Conclusions}. 
In this paper we have studied a QQ of the mass and coupling constant in the Sinh-Gordon model, in the special case of a large initial mass and zero initial interaction. We have seen that the ultraviolet regularization of this state is a non-trivial and physically relevant problem. This has led us to develop a systematic method to determine the expansion of the boundary state in the post-quench basis: this consists in solving integral equations for the excitation amplitudes $K_s$ which involve the finite-volume prescription of the exact Form Factors of the elementary field. Assuming that the boundary state is of the exponential form (\ref{sv}), we have obtained a first but 
quite accurate approximation of the solution by truncating the related Form Factor series. The proposed solution is used to derive the large time behaviour of observables.

The fact that the large energy behaviour of excitation amplitudes in the initial state is relevant for the calculation of physical observables at large times, means that models that are effectively equivalent as far as their ground state or thermal equilibrium properties are concerned, may not be equivalent out-of-equilibrium. 
More generally, we conclude that RG methods and concepts that are valid at equilibrium cannot always be applied directly to out-of-equilibrium problems.

It would be quite interesting to extend the analysis done in this paper to
other QQ protocols and determine the relevant amplitudes $K_s$ of the
corresponding boundary state from first principles. 

\

{\em Acknowledgements}. 
Spyros Sotiriadis acknowledges financial support by SISSA -- International School for Advanced Studies under the ``Young SISSA Scientists Research Projects'' scheme 2011-2012 and by the ERC under Starting Grant 279391 EDEQS. 
GT was partially supported by a Hungarian Academy of Sciences ``Momentum'' grant LP2012-50/2012. 
The work of GM is supported by the IRSES grants QICFT.
This work was also supported by the CNR-MTA joint project ``Nonperturbative field theory and strongly correlated systems". We would also like to thank the Max Planck Institute for the Physics of Complex Systems (Dresden, Germany) for
their hospitality during the international workshop QSOE'13.

\bibliography{QQshG_paper}

\begin{thebibliography}{46}%
\makeatletter
\providecommand \@ifxundefined [1]{%
 \@ifx{#1\undefined}
}%
\providecommand \@ifnum [1]{%
 \ifnum #1\expandafter \@firstoftwo
 \else \expandafter \@secondoftwo
 \fi
}%
\providecommand \@ifx [1]{%
 \ifx #1\expandafter \@firstoftwo
 \else \expandafter \@secondoftwo
 \fi
}%
\providecommand \natexlab [1]{#1}%
\providecommand \enquote  [1]{``#1''}%
\providecommand \bibnamefont  [1]{#1}%
\providecommand \bibfnamefont [1]{#1}%
\providecommand \citenamefont [1]{#1}%
\providecommand \href@noop [0]{\@secondoftwo}%
\providecommand \href [0]{\begingroup \@sanitize@url \@href}%
\providecommand \@href[1]{\@@startlink{#1}\@@href}%
\providecommand \@@href[1]{\endgroup#1\@@endlink}%
\providecommand \@sanitize@url [0]{\catcode `\\12\catcode `\$12\catcode
  `\&12\catcode `\#12\catcode `\^12\catcode `\_12\catcode `\%12\relax}%
\providecommand \@@startlink[1]{}%
\providecommand \@@endlink[0]{}%
\providecommand \url  [0]{\begingroup\@sanitize@url \@url }%
\providecommand \@url [1]{\endgroup\@href {#1}{\urlprefix }}%
\providecommand \urlprefix  [0]{URL }%
\providecommand \Eprint [0]{\href }%
\providecommand \doibase [0]{http://dx.doi.org/}%
\providecommand \selectlanguage [0]{\@gobble}%
\providecommand \bibinfo  [0]{\@secondoftwo}%
\providecommand \bibfield  [0]{\@secondoftwo}%
\providecommand \translation [1]{[#1]}%
\providecommand \BibitemOpen [0]{}%
\providecommand \bibitemStop [0]{}%
\providecommand \bibitemNoStop [0]{.\EOS\space}%
\providecommand \EOS [0]{\spacefactor3000\relax}%
\providecommand \BibitemShut  [1]{\csname bibitem#1\endcsname}%
\let\auto@bib@innerbib\@empty
\bibitem [{\citenamefont {Rammer}(2007)}]{Rammer}%
  \BibitemOpen
  \bibfield  {author} {\bibinfo {author} {\bibfnamefont {J.}~\bibnamefont
  {Rammer}},\ }\href@noop {} {\emph {\bibinfo {title} {Quantum Field Theory of
  Non-equilibrium States}}}\ (\bibinfo  {publisher} {Cambridge University
  Press},\ \bibinfo {year} {2007})\BibitemShut {NoStop}%
\bibitem [{\citenamefont {Calzetta}\ and\ \citenamefont {Hu}(2008)}]{Calzetta}%
  \BibitemOpen
  \bibfield  {author} {\bibinfo {author} {\bibfnamefont {E.}~\bibnamefont
  {Calzetta}}\ and\ \bibinfo {author} {\bibfnamefont {B.}~\bibnamefont {Hu}},\
  }\href@noop {} {\emph {\bibinfo {title} {Nonequilibrium Quantum Field
  Theory}}},\ Cambridge Monographs on Mathematical Physics\ (\bibinfo
  {publisher} {Cambridge University Press},\ \bibinfo {year}
  {2008})\BibitemShut {NoStop}%
\bibitem [{\citenamefont {Berges}(2004)}]{Berges}%
  \BibitemOpen
  \bibfield  {author} {\bibinfo {author} {\bibfnamefont {J.}~\bibnamefont
  {Berges}},\ }\href {http://dx.doi.org/10.1063/1.1843591} {\bibfield
  {journal} {\bibinfo  {journal} {AIP Conf. Proc.}\ }\textbf {\bibinfo {volume}
  {739}},\ \bibinfo {pages} {3} (\bibinfo {year} {2004})}\BibitemShut {NoStop}%
\bibitem [{\citenamefont {Mussardo}(2010)}]{Mussardo}%
  \BibitemOpen
  \bibfield  {author} {\bibinfo {author} {\bibfnamefont {G.}~\bibnamefont
  {Mussardo}},\ }\href@noop {} {\emph {\bibinfo {title} {Statistical Field
  Theory, An Introduction to Exactly Solved Models in Statistical Physics}}}\
  (\bibinfo  {publisher} {Oxford University Press, Oxford},\ \bibinfo {year}
  {2010})\BibitemShut {NoStop}%
\bibitem [{\citenamefont {Kinoshita}\ \emph {et~al.}(2006)\citenamefont
  {Kinoshita}, \citenamefont {Wenger},\ and\ \citenamefont
  {Weiss}}]{Kinoshita}%
  \BibitemOpen
  \bibfield  {author} {\bibinfo {author} {\bibfnamefont {T.}~\bibnamefont
  {Kinoshita}}, \bibinfo {author} {\bibfnamefont {T.}~\bibnamefont {Wenger}}, \
  and\ \bibinfo {author} {\bibfnamefont {D.~S.}\ \bibnamefont {Weiss}},\ }\href
  {http://dx.doi.org/10.1038/nature04693} {\bibfield  {journal} {\bibinfo
  {journal} {Nature}\ }\textbf {\bibinfo {volume} {440}},\ \bibinfo {pages}
  {900} (\bibinfo {year} {2006})}\BibitemShut {NoStop}%
\bibitem [{\citenamefont {Kormos}\ \emph {et~al.}(2009)\citenamefont {Kormos},
  \citenamefont {Mussardo},\ and\ \citenamefont {Trombettoni}}]{KMTPRL}%
  \BibitemOpen
  \bibfield  {author} {\bibinfo {author} {\bibfnamefont {M.}~\bibnamefont
  {Kormos}}, \bibinfo {author} {\bibfnamefont {G.}~\bibnamefont {Mussardo}}, \
  and\ \bibinfo {author} {\bibfnamefont {A.}~\bibnamefont {Trombettoni}},\
  }\href {\doibase 10.1103/PhysRevLett.103.210404} {\bibfield  {journal}
  {\bibinfo  {journal} {Phys. Rev. Lett.}\ }\textbf {\bibinfo {volume} {103}},\
  \bibinfo {pages} {210404} (\bibinfo {year} {2009})}\BibitemShut {NoStop}%
\bibitem [{\citenamefont {Kormos}\ \emph {et~al.}(2010)\citenamefont {Kormos},
  \citenamefont {Mussardo},\ and\ \citenamefont {Trombettoni}}]{KMT}%
  \BibitemOpen
  \bibfield  {author} {\bibinfo {author} {\bibfnamefont {M.}~\bibnamefont
  {Kormos}}, \bibinfo {author} {\bibfnamefont {G.}~\bibnamefont {Mussardo}}, \
  and\ \bibinfo {author} {\bibfnamefont {A.}~\bibnamefont {Trombettoni}},\
  }\href {\doibase 10.1103/PhysRevA.81.043606} {\bibfield  {journal} {\bibinfo
  {journal} {Phys. Rev. A}\ }\textbf {\bibinfo {volume} {81}},\ \bibinfo
  {pages} {043606} (\bibinfo {year} {2010})}\BibitemShut {NoStop}%
\bibitem [{\citenamefont {Rigol}\ \emph {et~al.}(2007)\citenamefont {Rigol},
  \citenamefont {Dunjko}, \citenamefont {Yurovsky},\ and\ \citenamefont
  {Olshanii}}]{Rigol}%
  \BibitemOpen
  \bibfield  {author} {\bibinfo {author} {\bibfnamefont {M.}~\bibnamefont
  {Rigol}}, \bibinfo {author} {\bibfnamefont {V.}~\bibnamefont {Dunjko}},
  \bibinfo {author} {\bibfnamefont {V.}~\bibnamefont {Yurovsky}}, \ and\
  \bibinfo {author} {\bibfnamefont {M.}~\bibnamefont {Olshanii}},\ }\href
  {\doibase 10.1103/PhysRevLett.98.050405} {\bibfield  {journal} {\bibinfo
  {journal} {Phys. Rev. Lett.}\ }\textbf {\bibinfo {volume} {98}},\ \bibinfo
  {eid} {050405} (\bibinfo {year} {2007})}\BibitemShut {NoStop}%
\bibitem [{\citenamefont {Cazalilla}(2006)}]{Cazalilla}%
  \BibitemOpen
  \bibfield  {author} {\bibinfo {author} {\bibfnamefont {M.~A.}\ \bibnamefont
  {Cazalilla}},\ }\href {\doibase 10.1103/PhysRevLett.97.156403} {\bibfield
  {journal} {\bibinfo  {journal} {Phys. Rev. Lett.;}\ }\textbf {\bibinfo
  {volume} {97}},\ \bibinfo {pages} {156403} (\bibinfo {year}
  {2006})}\BibitemShut {NoStop}%
\bibitem [{\citenamefont {Cazalilla}\ \emph {et~al.}(2012)\citenamefont
  {Cazalilla}, \citenamefont {Iucci},\ and\ \citenamefont {Chung}}]{Iucci}%
  \BibitemOpen
  \bibfield  {author} {\bibinfo {author} {\bibfnamefont {M.~A.}\ \bibnamefont
  {Cazalilla}}, \bibinfo {author} {\bibfnamefont {A.}~\bibnamefont {Iucci}}, \
  and\ \bibinfo {author} {\bibfnamefont {M.-C.}\ \bibnamefont {Chung}},\ }\href
  {\doibase 10.1103/PhysRevE.85.011133} {\bibfield  {journal} {\bibinfo
  {journal} {Phys. Rev. E}\ }\textbf {\bibinfo {volume} {85}},\ \bibinfo
  {pages} {011133} (\bibinfo {year} {2012})}\BibitemShut {NoStop}%
\bibitem [{\citenamefont {Iucci}\ and\ \citenamefont
  {Cazalilla}(2009)}]{Iucci2}%
  \BibitemOpen
  \bibfield  {author} {\bibinfo {author} {\bibfnamefont {A.}~\bibnamefont
  {Iucci}}\ and\ \bibinfo {author} {\bibfnamefont {M.~A.}\ \bibnamefont
  {Cazalilla}},\ }\href {\doibase 10.1103/PhysRevA.80.063619} {\bibfield
  {journal} {\bibinfo  {journal} {Phys. Rev. A}\ }\textbf {\bibinfo {volume}
  {80}},\ \bibinfo {pages} {063619} (\bibinfo {year} {2009})}\BibitemShut
  {NoStop}%
\bibitem [{\citenamefont {Calabrese}\ \emph {et~al.}(2011)\citenamefont
  {Calabrese}, \citenamefont {Essler},\ and\ \citenamefont {Fagotti}}]{Essler}%
  \BibitemOpen
  \bibfield  {author} {\bibinfo {author} {\bibfnamefont {P.}~\bibnamefont
  {Calabrese}}, \bibinfo {author} {\bibfnamefont {F.~H.~L.}\ \bibnamefont
  {Essler}}, \ and\ \bibinfo {author} {\bibfnamefont {M.}~\bibnamefont
  {Fagotti}},\ }\href {\doibase 10.1103/PhysRevLett.106.227203} {\bibfield
  {journal} {\bibinfo  {journal} {Phys. Rev. Lett.}\ }\textbf {\bibinfo
  {volume} {106}},\ \bibinfo {pages} {227203} (\bibinfo {year}
  {2011})}\BibitemShut {NoStop}%
\bibitem [{\citenamefont {Calabrese}\ \emph
  {et~al.}(2012{\natexlab{a}})\citenamefont {Calabrese}, \citenamefont
  {Essler},\ and\ \citenamefont {Fagotti}}]{CEF1}%
  \BibitemOpen
  \bibfield  {author} {\bibinfo {author} {\bibfnamefont {P.}~\bibnamefont
  {Calabrese}}, \bibinfo {author} {\bibfnamefont {F.~H.~L.}\ \bibnamefont
  {Essler}}, \ and\ \bibinfo {author} {\bibfnamefont {M.}~\bibnamefont
  {Fagotti}},\ }\href {http://stacks.iop.org/1742-5468/2012/i=07/a=P07016}
  {\bibfield  {journal} {\bibinfo  {journal} {J. Stat. Mech.}\ }\textbf
  {\bibinfo {volume} {2012}},\ \bibinfo {pages} {P07016} (\bibinfo {year}
  {2012}{\natexlab{a}})}\BibitemShut {NoStop}%
\bibitem [{\citenamefont {Calabrese}\ \emph
  {et~al.}(2012{\natexlab{b}})\citenamefont {Calabrese}, \citenamefont
  {Essler},\ and\ \citenamefont {Fagotti}}]{CEF2}%
  \BibitemOpen
  \bibfield  {author} {\bibinfo {author} {\bibfnamefont {P.}~\bibnamefont
  {Calabrese}}, \bibinfo {author} {\bibfnamefont {F.~H.~L.}\ \bibnamefont
  {Essler}}, \ and\ \bibinfo {author} {\bibfnamefont {M.}~\bibnamefont
  {Fagotti}},\ }\href {http://stacks.iop.org/1742-5468/2012/i=07/a=P07022}
  {\bibfield  {journal} {\bibinfo  {journal} {J. Stat. Mech.}\ }\textbf
  {\bibinfo {volume} {2012}},\ \bibinfo {pages} {P07022} (\bibinfo {year}
  {2012}{\natexlab{b}})}\BibitemShut {NoStop}%
\bibitem [{\citenamefont {Schuricht}\ and\ \citenamefont {Essler}(2012)}]{SE}%
  \BibitemOpen
  \bibfield  {author} {\bibinfo {author} {\bibfnamefont {D.}~\bibnamefont
  {Schuricht}}\ and\ \bibinfo {author} {\bibfnamefont {F.~H.~L.}\ \bibnamefont
  {Essler}},\ }\href {http://stacks.iop.org/1742-5468/2012/i=04/a=P04017}
  {\bibfield  {journal} {\bibinfo  {journal} {J. Stat. Mech.}\ }\textbf
  {\bibinfo {volume} {2012}},\ \bibinfo {pages} {P04017} (\bibinfo {year}
  {2012})}\BibitemShut {NoStop}%
\bibitem [{\citenamefont {Collura}\ \emph
  {et~al.}(2013{\natexlab{a}})\citenamefont {Collura}, \citenamefont
  {Sotiriadis},\ and\ \citenamefont {Calabrese}}]{CSC13a}%
  \BibitemOpen
  \bibfield  {author} {\bibinfo {author} {\bibfnamefont {M.}~\bibnamefont
  {Collura}}, \bibinfo {author} {\bibfnamefont {S.}~\bibnamefont {Sotiriadis}},
  \ and\ \bibinfo {author} {\bibfnamefont {P.}~\bibnamefont {Calabrese}},\
  }\href {\doibase 10.1103/PhysRevLett.110.245301} {\bibfield  {journal}
  {\bibinfo  {journal} {Phys. Rev. Lett.}\ }\textbf {\bibinfo {volume} {110}},\
  \bibinfo {pages} {245301} (\bibinfo {year} {2013}{\natexlab{a}})}\BibitemShut
  {NoStop}%
\bibitem [{\citenamefont {Collura}\ \emph
  {et~al.}(2013{\natexlab{b}})\citenamefont {Collura}, \citenamefont
  {Sotiriadis},\ and\ \citenamefont {Calabrese}}]{CSC13b}%
  \BibitemOpen
  \bibfield  {author} {\bibinfo {author} {\bibfnamefont {M.}~\bibnamefont
  {Collura}}, \bibinfo {author} {\bibfnamefont {S.}~\bibnamefont {Sotiriadis}},
  \ and\ \bibinfo {author} {\bibfnamefont {P.}~\bibnamefont {Calabrese}},\
  }\href {http://stacks.iop.org/1742-5468/2013/i=09/a=P09025} {\bibfield
  {journal} {\bibinfo  {journal} {J. Stat. Mech.}\ }\textbf {\bibinfo {volume}
  {2013}},\ \bibinfo {pages} {P09025} (\bibinfo {year}
  {2013}{\natexlab{b}})}\BibitemShut {NoStop}%
\bibitem [{\citenamefont {Kormos}\ \emph
  {et~al.}(2013{\natexlab{a}})\citenamefont {Kormos}, \citenamefont {Collura},\
  and\ \citenamefont {Calabrese}}]{Kormos13b}%
  \BibitemOpen
  \bibfield  {author} {\bibinfo {author} {\bibfnamefont {M.}~\bibnamefont
  {Kormos}}, \bibinfo {author} {\bibfnamefont {M.}~\bibnamefont {Collura}}, \
  and\ \bibinfo {author} {\bibfnamefont {P.}~\bibnamefont {Calabrese}},\ }\href
  {http://arxiv.org/abs/1307.2142} {\bibfield  {journal} {\bibinfo  {journal}
  {arXiv:1307.2142}\ } (\bibinfo {year} {2013}{\natexlab{a}})}\BibitemShut
  {NoStop}%
\bibitem [{\citenamefont {Goldstein}\ and\ \citenamefont
  {Andrei}(2013)}]{Andrei13c}%
  \BibitemOpen
  \bibfield  {author} {\bibinfo {author} {\bibfnamefont {G.}~\bibnamefont
  {Goldstein}}\ and\ \bibinfo {author} {\bibfnamefont {N.}~\bibnamefont
  {Andrei}},\ }\href {http://arxiv.org/abs/1309.3471} {\bibfield  {journal}
  {\bibinfo  {journal} {arXiv:1309.3471}\ } (\bibinfo {year}
  {2013})}\BibitemShut {NoStop}%
\bibitem [{\citenamefont {Fioretto}\ and\ \citenamefont
  {Mussardo}(2010)}]{Fioretto}%
  \BibitemOpen
  \bibfield  {author} {\bibinfo {author} {\bibfnamefont {D.}~\bibnamefont
  {Fioretto}}\ and\ \bibinfo {author} {\bibfnamefont {G.}~\bibnamefont
  {Mussardo}},\ }\href {http://stacks.iop.org/1367-2630/12/i=5/a=055015}
  {\bibfield  {journal} {\bibinfo  {journal} {New J. Phys.}\ }\textbf {\bibinfo
  {volume} {12}},\ \bibinfo {pages} {055015} (\bibinfo {year}
  {2010})}\BibitemShut {NoStop}%
\bibitem [{\citenamefont {Fagotti}\ and\ \citenamefont {Essler}(2013)}]{FE13}%
  \BibitemOpen
  \bibfield  {author} {\bibinfo {author} {\bibfnamefont {M.}~\bibnamefont
  {Fagotti}}\ and\ \bibinfo {author} {\bibfnamefont {F.~H.~L.}\ \bibnamefont
  {Essler}},\ }\href {http://stacks.iop.org/1742-5468/2013/i=07/a=P07012}
  {\bibfield  {journal} {\bibinfo  {journal} {J. Stat. Mech.}\ }\textbf
  {\bibinfo {volume} {2013}},\ \bibinfo {pages} {P07012} (\bibinfo {year}
  {2013})}\BibitemShut {NoStop}%
\bibitem [{\citenamefont {Caux}\ and\ \citenamefont {Essler}(2013)}]{CE}%
  \BibitemOpen
  \bibfield  {author} {\bibinfo {author} {\bibfnamefont {J.-S.}\ \bibnamefont
  {Caux}}\ and\ \bibinfo {author} {\bibfnamefont {F.~H.~L.}\ \bibnamefont
  {Essler}},\ }\href {\doibase 10.1103/PhysRevLett.110.257203} {\bibfield
  {journal} {\bibinfo  {journal} {Phys. Rev. Lett.}\ }\textbf {\bibinfo
  {volume} {110}},\ \bibinfo {pages} {257203} (\bibinfo {year}
  {2013})}\BibitemShut {NoStop}%
\bibitem [{\citenamefont {Pozsgay}(2013{\natexlab{a}})}]{Pozsgay13a}%
  \BibitemOpen
  \bibfield  {author} {\bibinfo {author} {\bibfnamefont {B.}~\bibnamefont
  {Pozsgay}},\ }\href {http://stacks.iop.org/1742-5468/2013/i=07/a=P07003}
  {\bibfield  {journal} {\bibinfo  {journal} {J. Stat. Mech.}\ }\textbf
  {\bibinfo {volume} {2013}},\ \bibinfo {pages} {P07003} (\bibinfo {year}
  {2013}{\natexlab{a}})}\BibitemShut {NoStop}%
\bibitem [{\citenamefont {Kormos}\ \emph
  {et~al.}(2013{\natexlab{b}})\citenamefont {Kormos}, \citenamefont {Shashi},
  \citenamefont {Chou}, \citenamefont {Caux},\ and\ \citenamefont
  {Imambekov}}]{Kormos13}%
  \BibitemOpen
  \bibfield  {author} {\bibinfo {author} {\bibfnamefont {M.}~\bibnamefont
  {Kormos}}, \bibinfo {author} {\bibfnamefont {A.}~\bibnamefont {Shashi}},
  \bibinfo {author} {\bibfnamefont {Y.-Z.}\ \bibnamefont {Chou}}, \bibinfo
  {author} {\bibfnamefont {J.-S.}\ \bibnamefont {Caux}}, \ and\ \bibinfo
  {author} {\bibfnamefont {A.}~\bibnamefont {Imambekov}},\ }\href
  {http://arxiv.org/abs/1305.7202} {\bibfield  {journal} {\bibinfo  {journal}
  {arXiv:1305.7202}\ } (\bibinfo {year} {2013}{\natexlab{b}})}\BibitemShut
  {NoStop}%
\bibitem [{\citenamefont {Rajabpour}\ and\ \citenamefont
  {Sotiriadis}(2013)}]{RS13}%
  \BibitemOpen
  \bibfield  {author} {\bibinfo {author} {\bibfnamefont {M.~A.}\ \bibnamefont
  {Rajabpour}}\ and\ \bibinfo {author} {\bibfnamefont {S.}~\bibnamefont
  {Sotiriadis}},\ }\href {http://arxiv.org/abs/1307.7697} {\bibfield  {journal}
  {\bibinfo  {journal} {arXiv:1307.7697}\ } (\bibinfo {year}
  {2013})}\BibitemShut {NoStop}%
\bibitem [{\citenamefont {Nardis}\ \emph {et~al.}(2013)\citenamefont {Nardis},
  \citenamefont {Wouters}, \citenamefont {Brockmann},\ and\ \citenamefont
  {Caux}}]{Caux13}%
  \BibitemOpen
  \bibfield  {author} {\bibinfo {author} {\bibfnamefont {J.~D.}\ \bibnamefont
  {Nardis}}, \bibinfo {author} {\bibfnamefont {B.}~\bibnamefont {Wouters}},
  \bibinfo {author} {\bibfnamefont {M.}~\bibnamefont {Brockmann}}, \ and\
  \bibinfo {author} {\bibfnamefont {J.-S.}\ \bibnamefont {Caux}},\ }\href
  {http://arxiv.org/abs/1308.4310} {\bibfield  {journal} {\bibinfo  {journal}
  {arXiv:1308.4310}\ } (\bibinfo {year} {2013})}\BibitemShut {NoStop}%
\bibitem [{\citenamefont {Pozsgay}(2013{\natexlab{b}})}]{Pozsgay13b}%
  \BibitemOpen
  \bibfield  {author} {\bibinfo {author} {\bibfnamefont {B.}~\bibnamefont
  {Pozsgay}},\ }\href {http://arxiv.org/abs/1309.4593} {\bibfield  {journal}
  {\bibinfo  {journal} {arXiv:1309.4593}\ } (\bibinfo {year}
  {2013}{\natexlab{b}})}\BibitemShut {NoStop}%
\bibitem [{\citenamefont {Mussardo}(2013)}]{Muss13}%
  \BibitemOpen
  \bibfield  {author} {\bibinfo {author} {\bibfnamefont {G.}~\bibnamefont
  {Mussardo}},\ }\href {\doibase 10.1103/PhysRevLett.111.100401} {\bibfield
  {journal} {\bibinfo  {journal} {Phys. Rev. Lett.}\ }\textbf {\bibinfo
  {volume} {111}},\ \bibinfo {pages} {100401} (\bibinfo {year}
  {2013})}\BibitemShut {NoStop}%
\bibitem [{\citenamefont {Iyer}\ \emph {et~al.}(2013)\citenamefont {Iyer},
  \citenamefont {Guan},\ and\ \citenamefont {Andrei}}]{Andrei13a}%
  \BibitemOpen
  \bibfield  {author} {\bibinfo {author} {\bibfnamefont {D.}~\bibnamefont
  {Iyer}}, \bibinfo {author} {\bibfnamefont {H.}~\bibnamefont {Guan}}, \ and\
  \bibinfo {author} {\bibfnamefont {N.}~\bibnamefont {Andrei}},\ }\href
  {\doibase 10.1103/PhysRevA.87.053628} {\bibfield  {journal} {\bibinfo
  {journal} {Phys. Rev. A}\ }\textbf {\bibinfo {volume} {87}},\ \bibinfo
  {pages} {053628} (\bibinfo {year} {2013})}\BibitemShut {NoStop}%
\bibitem [{\citenamefont {Liu}\ and\ \citenamefont {Andrei}(2013)}]{Andrei13b}%
  \BibitemOpen
  \bibfield  {author} {\bibinfo {author} {\bibfnamefont {W.}~\bibnamefont
  {Liu}}\ and\ \bibinfo {author} {\bibfnamefont {N.}~\bibnamefont {Andrei}},\
  }\href {http://arxiv.org/abs/1311.1118} {\bibfield  {journal} {\bibinfo
  {journal} {arXiv:1311.1118}\ } (\bibinfo {year} {2013})}\BibitemShut
  {NoStop}%
\bibitem [{\citenamefont {Calabrese}\ and\ \citenamefont {Cardy}(2006)}]{CC1}%
  \BibitemOpen
  \bibfield  {author} {\bibinfo {author} {\bibfnamefont {P.}~\bibnamefont
  {Calabrese}}\ and\ \bibinfo {author} {\bibfnamefont {J.}~\bibnamefont
  {Cardy}},\ }\href {\doibase 10.1103/PhysRevLett.96.136801} {\bibfield
  {journal} {\bibinfo  {journal} {Phys. Rev. Lett.}\ }\textbf {\bibinfo
  {volume} {96}},\ \bibinfo {eid} {136801} (\bibinfo {year}
  {2006})}\BibitemShut {NoStop}%
\bibitem [{\citenamefont {Pozsgay}(2011)}]{Pozsgay}%
  \BibitemOpen
  \bibfield  {author} {\bibinfo {author} {\bibfnamefont {B.}~\bibnamefont
  {Pozsgay}},\ }\href {http://stacks.iop.org/1742-5468/2011/i=01/a=P01011}
  {\bibfield  {journal} {\bibinfo  {journal} {J. Stat. Mech.}\ }\textbf
  {\bibinfo {volume} {2011}},\ \bibinfo {pages} {P01011} (\bibinfo {year}
  {2011})}\BibitemShut {NoStop}%
\bibitem [{\citenamefont {Ghoshal}\ and\ \citenamefont
  {Zamolodchikov}(1994)}]{Ghoshal}%
  \BibitemOpen
  \bibfield  {author} {\bibinfo {author} {\bibfnamefont {S.}~\bibnamefont
  {Ghoshal}}\ and\ \bibinfo {author} {\bibfnamefont {A.}~\bibnamefont
  {Zamolodchikov}},\ }\href {http://dx.doi.org/10.1142/S0217751X94001941}
  {\bibfield  {journal} {\bibinfo  {journal} {Int. J. Mod. Phys. A}\ }\textbf
  {\bibinfo {volume} {9}},\ \bibinfo {pages} {3841} (\bibinfo {year}
  {1994})}\BibitemShut {NoStop}%
\bibitem [{\citenamefont {Calabrese}\ and\ \citenamefont {Cardy}(2007)}]{CC2}%
  \BibitemOpen
  \bibfield  {author} {\bibinfo {author} {\bibfnamefont {P.}~\bibnamefont
  {Calabrese}}\ and\ \bibinfo {author} {\bibfnamefont {J.}~\bibnamefont
  {Cardy}},\ }\href {http://stacks.iop.org/1742-5468/2007/i=06/a=P06008}
  {\bibfield  {journal} {\bibinfo  {journal} {J. Stat. Mech.}\ }\textbf
  {\bibinfo {volume} {2007}},\ \bibinfo {pages} {P06008} (\bibinfo {year}
  {2007})}\BibitemShut {NoStop}%
\bibitem [{\citenamefont {Cardy}(2012{\natexlab{a}})}]{Cardy12a}%
  \BibitemOpen
  \bibfield  {author} {\bibinfo {author} {\bibfnamefont {J.}~\bibnamefont
  {Cardy}},\ }\href {http://www.ggi.fi.infn.it/talks/talk2326.pdf} {\enquote
  {\bibinfo {title} {{Quantum Quenches in Perturbed Conformal Field
  Theories}},}\ } (\bibinfo {year} {2012}{\natexlab{a}}),\ \bibinfo {note}
  {talk at GGI workshop: New quantum states of matter in and out of
  equilibrium}\BibitemShut {NoStop}%
\bibitem [{\citenamefont {Cardy}(2012{\natexlab{b}})}]{Cardy12b}%
  \BibitemOpen
  \bibfield  {author} {\bibinfo {author} {\bibfnamefont {J.}~\bibnamefont
  {Cardy}},\ }\href {http://online.kitp.ucsb.edu/online/qdynamics-c12/cardy/}
  {\enquote {\bibinfo {title} {{Quantum Quench in a Conformal Field Theory From
  a General Short-Range State}},}\ } (\bibinfo {year} {2012}{\natexlab{b}}),\
  \bibinfo {note} {talk at KITP Conference: Dynamics and Thermodynamics in
  Isolated Quantum Systems}\BibitemShut {NoStop}%
\bibitem [{\citenamefont {Fring}\ \emph {et~al.}(1993)\citenamefont {Fring},
  \citenamefont {Mussardo},\ and\ \citenamefont {Simonetti}}]{Fring:1992pt}%
  \BibitemOpen
  \bibfield  {author} {\bibinfo {author} {\bibfnamefont {A.}~\bibnamefont
  {Fring}}, \bibinfo {author} {\bibfnamefont {G.}~\bibnamefont {Mussardo}}, \
  and\ \bibinfo {author} {\bibfnamefont {P.}~\bibnamefont {Simonetti}},\ }\href
  {\doibase http://dx.doi.org/10.1016/0550-3213(93)90252-K} {\bibfield
  {journal} {\bibinfo  {journal} {Nuclear Physics B}\ }\textbf {\bibinfo
  {volume} {393}},\ \bibinfo {pages} {413 } (\bibinfo {year}
  {1993})}\BibitemShut {NoStop}%
\bibitem [{\citenamefont {Koubek}\ and\ \citenamefont {Mussardo}(1993)}]{KM}%
  \BibitemOpen
  \bibfield  {author} {\bibinfo {author} {\bibfnamefont {A.}~\bibnamefont
  {Koubek}}\ and\ \bibinfo {author} {\bibfnamefont {G.}~\bibnamefont
  {Mussardo}},\ }\href {http://dx.doi.org/10.1016/0370-2693(93)90554-U}
  {\bibfield  {journal} {\bibinfo  {journal} {Phys. Lett. B}\ }\textbf
  {\bibinfo {volume} {311}},\ \bibinfo {pages} {193 } (\bibinfo {year}
  {1993})}\BibitemShut {NoStop}%
\bibitem [{\citenamefont {Sotiriadis}\ \emph {et~al.}(2012)\citenamefont
  {Sotiriadis}, \citenamefont {Fioretto},\ and\ \citenamefont
  {Mussardo}}]{SFM}%
  \BibitemOpen
  \bibfield  {author} {\bibinfo {author} {\bibfnamefont {S.}~\bibnamefont
  {Sotiriadis}}, \bibinfo {author} {\bibfnamefont {D.}~\bibnamefont
  {Fioretto}}, \ and\ \bibinfo {author} {\bibfnamefont {G.}~\bibnamefont
  {Mussardo}},\ }\href {http://stacks.iop.org/1742-5468/2012/i=02/a=P02017}
  {\bibfield  {journal} {\bibinfo  {journal} {J. Stat. Mech.}\ }\textbf
  {\bibinfo {volume} {2012}},\ \bibinfo {pages} {P02017} (\bibinfo {year}
  {2012})}\BibitemShut {NoStop}%
\bibitem [{\citenamefont {Ghoshal}(1994)}]{Ghoshal2}%
  \BibitemOpen
  \bibfield  {author} {\bibinfo {author} {\bibfnamefont {S.}~\bibnamefont
  {Ghoshal}},\ }\href {\doibase 10.1142/S0217751X94001941} {\bibfield
  {journal} {\bibinfo  {journal} {Int. J. Mod. Phys. A}\ }\textbf {\bibinfo
  {volume} {09}},\ \bibinfo {pages} {4801} (\bibinfo {year}
  {1994})}\BibitemShut {NoStop}%
\bibitem [{\citenamefont {Smirnov}(1992)}]{Smirnov}%
  \BibitemOpen
  \bibfield  {author} {\bibinfo {author} {\bibfnamefont {F.}~\bibnamefont
  {Smirnov}},\ }\href@noop {} {\emph {\bibinfo {title} {Form Factors in
  Completely Integrable Models of Quantum Field Theory}}},\ Advanced series in
  mathematical physics\ (\bibinfo  {publisher} {World Scientific},\ \bibinfo
  {year} {1992})\BibitemShut {NoStop}%
\bibitem [{\citenamefont {Pozsgay}\ and\ \citenamefont
  {Takacs}(2010)}]{PozsgayTakacs}%
  \BibitemOpen
  \bibfield  {author} {\bibinfo {author} {\bibfnamefont {B.}~\bibnamefont
  {Pozsgay}}\ and\ \bibinfo {author} {\bibfnamefont {G.}~\bibnamefont
  {Takacs}},\ }\href {http://stacks.iop.org/1742-5468/2010/i=11/a=P11012}
  {\bibfield  {journal} {\bibinfo  {journal} {J. Stat. Mech.}\ }\textbf
  {\bibinfo {volume} {2010}},\ \bibinfo {pages} {P11012} (\bibinfo {year}
  {2010})}\BibitemShut {NoStop}%
\bibitem [{\citenamefont {Kormos}\ and\ \citenamefont
  {Pozsgay}(2010)}]{KormosPozsgay}%
  \BibitemOpen
  \bibfield  {author} {\bibinfo {author} {\bibfnamefont {M.}~\bibnamefont
  {Kormos}}\ and\ \bibinfo {author} {\bibfnamefont {B.}~\bibnamefont
  {Pozsgay}},\ }\href {\doibase 10.1007/JHEP04(2010)112} {\bibfield  {journal}
  {\bibinfo  {journal} {JHEP}\ }\textbf {\bibinfo {volume} {2010}},\ \bibinfo
  {pages} {1} (\bibinfo {year} {2010})}\BibitemShut {NoStop}%
\bibitem [{\citenamefont {Pozsgay}\ and\ \citenamefont
  {Takacs}(2008{\natexlab{a}})}]{Pozsgay:2007kn}%
  \BibitemOpen
  \bibfield  {author} {\bibinfo {author} {\bibfnamefont {B.}~\bibnamefont
  {Pozsgay}}\ and\ \bibinfo {author} {\bibfnamefont {G.}~\bibnamefont
  {Takacs}},\ }\href {\doibase 10.1016/j.nuclphysb.2007.06.027} {\bibfield
  {journal} {\bibinfo  {journal} {Nucl. Phys.}\ }\textbf {\bibinfo {volume}
  {B788}},\ \bibinfo {pages} {167} (\bibinfo {year} {2008}{\natexlab{a}})},\
  \Eprint {http://arxiv.org/abs/0706.1445} {arXiv:0706.1445 [hep-th]}
  \BibitemShut {NoStop}%
\bibitem [{\citenamefont {Pozsgay}\ and\ \citenamefont
  {Takacs}(2008{\natexlab{b}})}]{Pozsgay:2007gx}%
  \BibitemOpen
  \bibfield  {author} {\bibinfo {author} {\bibfnamefont {B.}~\bibnamefont
  {Pozsgay}}\ and\ \bibinfo {author} {\bibfnamefont {G.}~\bibnamefont
  {Takacs}},\ }\href {\doibase 10.1016/j.nuclphysb.2007.07.008} {\bibfield
  {journal} {\bibinfo  {journal} {Nucl.Phys.}\ }\textbf {\bibinfo {volume}
  {B788}},\ \bibinfo {pages} {209} (\bibinfo {year} {2008}{\natexlab{b}})},\
  \Eprint {http://arxiv.org/abs/0706.3605} {arXiv:0706.3605 [hep-th]}
  \BibitemShut {NoStop}%
\bibitem [{\citenamefont {Sotiriadis}\ \emph {et~al.}()\citenamefont
  {Sotiriadis}, \citenamefont {Takacs},\ and\ \citenamefont
  {Mussardo}}]{long_version}%
  \BibitemOpen
  \bibfield  {author} {\bibinfo {author} {\bibfnamefont {S.}~\bibnamefont
  {Sotiriadis}}, \bibinfo {author} {\bibfnamefont {G.}~\bibnamefont {Takacs}},
  \ and\ \bibinfo {author} {\bibfnamefont {G.}~\bibnamefont {Mussardo}},\
  }\href@noop {} {\emph {\bibinfo {title} {(in preparation)}}}\BibitemShut
  {NoStop}%
\end{thebibliography}%


\begin{thebibliography}{1}


\bibitem{AFZ} A.E. Arinshtein, V.A. Fateev and A.B. Zamolodchikov,
{\em Phys. Lett.} \textbf{87B} (1979), 389 

\bibitem{Karowski}B. Berg, M. Karowski, P. Weisz, {\em Phys. Rev.}
\textbf{D19} (1979), 2477; M. Karowski, P. Weisz, {\em Nucl. Phys.}
\textbf{B139} (1978), 445; M. Karowski, {\em Phys. Rep.} \textbf{49}
(1979), 229 

\bibitem{Smirnov} F.A. Smirnov, Form Factors in Completely Integrable
Models of Quantum Field Theory (World Scientific) 1992, and references
therein. 

\bibitem{FMS} A.Fring, G.Mussardo, P.Simonetti, {\em Nucl. Phys.}
\textbf{B 393}, (1993), 413 

\bibitem{KM} A. Koubek and G. Mussardo, {\em Phys. Lett.} \textbf{B
311} (1993), 193.

\bibitem{Pozsgay:2007kn} B.~Pozsgay and G.~Takacs, ``{Form factors
in finite volume I: form factor bootstrap and truncated conformal
space},'' {{\em Nucl. Phys.} \textbf{B788} (2008) 167--208}. 

\bibitem{Pozsgay:2007gx} B.~Pozsgay and G.~Takacs, ``{Form factors
in finite volume. II. Disconnected terms and finite temperature correlators},''
{{\em Nucl.Phys.} \textbf{B788} (2008) 209--251}. 

\bibitem{Kormos:2010}M. Kormos and B. Pozsgay: ``One-Point Functions
in Massive Integrable QFT with Boundaries,'' JHEP 1004 (2010) 112,
arXiv:1002.2783 {[}hep-th{]}. 

\bibitem{Pozsgay:2010cr} B.~Pozsgay and G.~Takacs, ``{Form factor
expansion for thermal correlators},'' {{\em J.Stat.Mech.} \textbf{1011}
(2010) P11012}.

\bibitem{CC1}
P.~Calabrese and J.~Cardy, 
Phys. Rev. Lett. \textbf {96} 136801 (2006).
\bibitem{CC2}
P.~Calabrese and J.~Cardy, 
J. Stat. Mech. \textbf{2007} P06008 (2007).


 \end{thebibliography}


\onecolumngrid
\newpage

\begin{center}
{\Large{\bf Supplementary Material}}
\end{center}

\section{Form Factors of the Sinh-Gordon model}

The Sinh-Gordon theory is defined by the action 
\begin{equation}
{\cal S}\,=\,\int d^{2}x\left[\frac{1}{2}(\partial_{\mu}\phi)^{2}-\frac{m_{0}^{2}}{g^{2}}\,\cosh\, g\phi(x)\,\right]\,\,.\label{Lagrangian}
\end{equation}
It is the simplest example of Lagrangian integrable field theory and
it is invariant under the $Z_{2}$ symmetry $\phi\rightarrow-\phi$.
Parameterising the dispersion relations in terms of the rapidity $\beta$,
$p^{0}\,=\, m\cosh\beta\,\,\,,p^{1}\,=\, m\sinh\beta\,\,\,$, its
two-particle $S$-matrix is given by \cite{AFZ} 
\begin{equation}
S(\beta,B)\,=\,\frac{\tanh\frac{1}{2}(\beta-i\frac{\pi B}{2})}{\tanh\frac{1}{2}(\beta+i\frac{\pi B}{2})}\,\,,\label{smatrix}
\end{equation}
where $\beta=\beta_{1}-\beta_{2}$ and $B$ is the so-called renormalized
coupling constant 
\begin{equation}
B(g)\,=\,\frac{2g^{2}}{8\pi+g^{2}}\sm.
\end{equation}
For real values of $g$ the $S$-matrix has no poles in the physical
sheet and hence there are no bound states.

The form factors (FF) $F_{n}^{{\cal O}}$ of the Sinh-Gordon model
are matrix elements of a generic local operator ${\cal O}$ between
the vacuum and a set of $n$ particle asymptotic states 
\begin{equation}
F_{n}^{{\cal O}}(\beta_{1},\beta_{2},\ldots,\beta_{n})\,=\,\langle0\mid{\cal O}(0,0)\mid\beta_{1},\beta_{2},\ldots,\beta_{n}\rangle_{in}\,\,\,.\label{FF}
\end{equation}
Based on general properties of a QFT (as unitarity, analyticity and
locality), the form factor bootstrap approach leads to a system of
linear and recursive equations for the matrix elements $F_{n}^{{\cal O}}$
\cite{Karowski,Smirnov} 
\begin{eqnarray}
F_{n}(\beta_{1},\dots,\beta_{i},\beta_{i+1},\dots,\beta_{n}) & = & F_{n}(\beta_{1},\dots,\beta_{i+1},\beta_{i},\dots,\beta_{n})\, S(\beta_{i}-\beta_{i+1})\,\,,\nonumber \\
F_{n}(\beta_{1}+2\pi i,\dots,\beta_{n-1},\beta_{n}) & = & F_{n}(\beta_{2},\ldots,\beta_{n-1},\beta_{n},\beta_{1})\,\,,\label{axioms}\\
-i\lim_{\tilde{\beta}\rightarrow\beta}(\tilde{\beta}-\beta)F_{n+2}(\tilde{\beta}+i\pi,\beta,\ldots,\beta_{n}) & = & \left(1-\prod_{i=1}^{n}S(\beta-\beta_{i})\right)\, F_{n}(\beta_{1},\ldots,\beta_{n})\,\,\,.\nonumber 
\end{eqnarray}
For an operator ${\cal O}(x)$ of spin $s$, relativistic invariance
implies 
\begin{equation}
F_{n}^{{\cal O}}(\beta_{1}+\Lambda,\beta_{2}+\Lambda,\ldots,\beta_{n}+\Lambda)\,=\, e^{s\Lambda}\, F_{n}^{{\cal O}}(\beta_{1},\beta_{2},\ldots,\beta_{n})\,\,.\label{asymp2}
\end{equation}
In the Sinh-Gordon model a convenient parameterization of the $n$-particle
FF, which takes into account their kinematical poles, is given by
\cite{FMS} 
\begin{equation}
F_{n}(\beta_{1},\ldots,\beta_{n})\,=\, H_{n}\, Q_{n}(x_{1},\ldots,x_{n})\,\prod_{i<j}\frac{F_{{\rm min}}(\beta_{ij})}{(x_{i}+x_{j})}\sb,\label{para}
\end{equation}
where $x_{i}\equiv e^{\beta_{i}}$ and $\beta_{ij}=\beta_{i}-\beta_{j}$.
$F_{{\rm min}}(\beta)$ is an analytic function given by 
\begin{equation}
F_{{\rm min}}(\beta,B)\,=\,\prod_{k=0}^{\infty}\left|\frac{\Gamma\left(k+\frac{3}{2}+\frac{i\hat{\beta}}{2\pi}\right)\Gamma\left(k+\frac{1}{2}+\frac{B}{4}+\frac{i\hat{\beta}}{2\pi}\right)\Gamma\left(k+1-\frac{B}{4}+\frac{i\hat{\beta}}{2\pi}\right)}{\Gamma\left(k+\frac{1}{2}+\frac{i\hat{\beta}}{2\pi}\right)\Gamma\left(k+\frac{3}{2}-\frac{B}{4}+\frac{i\hat{\beta}}{2\pi}\right)\Gamma\left(k+1+\frac{B}{4}+\frac{i\hat{\beta}}{2\pi}\right)}\right|^{2}\sa,
\end{equation}
that satisfies the functional equations 
\begin{equation}
\begin{array}{ccl}
F_{{\rm min}}(\beta) & = & F_{{\rm min}}(-\beta)\, S(\beta,B)\,\,,\\
F_{{\rm min}}(i\pi-\beta) & = & F_{{\rm min}}(i\pi+\beta)\,\,,\\
F_{{\rm min}}(i\pi+\beta,B)F_{{\rm min}}(\beta,B) & = & \frac{\sinh\beta}{\sinh\beta+\sinh\frac{i\pi B}{2}}\,\,\,.
\end{array}\label{Watson2}
\end{equation}
It has a simple zero at the threshold $\beta=0$ (since $S(0,B)=-1$)
and no poles in the physical strip $0\leq{\rm Im}\,\beta\leq\pi$,
with an asymptotic behaviour given by 
\begin{equation}
\lim_{\beta\rightarrow\infty}F_{{\rm min}}(\beta,B)=1\,\,.\label{limitefmin}
\end{equation}
$H_{n}$ are normalization constants, which can be conveniently chosen
as 
\begin{equation}
H_{2n+1}=H_{1}\mu^{2n}\sb,\sb H_{2n}=H_{2}\mu^{2n-2}\,\,\,,
\end{equation}
with 
\begin{equation}
\mu\equiv\left(\frac{4\sin(\pi B/2)}{F_{min}(i\pi,B)}\right)^{\frac{1}{2}}
\end{equation}
and $H_{1}$, $H_{2}$ are two independent parameters. Finally, the
functions $Q_{n}(x_{1},\dots,x_{n})$ are symmetric polynomials in
the variables $x_{i}$, which have to be fixed by the recursion equations
satisfied by the form factors. For FF of spinless operators, their
total degree is equal to $n(n-1)/2$. On the other hand, the partial
degree of $Q_{n}$ in each variable $x_{i}$ is fixed by the nature
and the asymptotic behaviour of the operator ${\cal O}$ which is
considered.

Exploiting the parameterization (\ref{para}) together with the functional
equations (\ref{axioms}) and (\ref{Watson2}), the polynomials $Q_{n}(x_{1},\ldots,x_{n})$
have to satisfy the recursive equations \cite{FMS} 
\begin{equation}
(-)^{n}\, Q_{n+2}(-x,x,x_{1},\ldots,x_{n})\,=\, xD_{n}(x,x_{1},x_{2},\ldots,x_{n})\, Q_{n}(x_{1},x_{2},\ldots,x_{n})\label{rec}
\end{equation}
with 
\begin{equation}
D_{n}(x,x_{1},\dots,x_{n})=\sum_{k=1}^{n}\sum_{m=1,odd}^{k}[m]\, x^{2(n-k)+m}\sigma_{k}^{(n)}\sigma_{k-m}^{(n)}(-1)^{k+1}\sb.\label{D_n}
\end{equation}
We have introduced the symbol $[n]$ defined by 
\begin{equation}
[n]\equiv\frac{\sin(n\frac{B}{2})}{\sin\frac{B}{2}}
\end{equation}
and the \emph{elementary symmetric polynomials} $\sigma_{k}^{(n)}(x_{1},\dots,x_{n})$,
given by the generating function 
\begin{equation}
\prod_{i=1}^{n}(x+x_{i})\,=\,\sum_{k=0}^{n}x^{n-k}\,\sigma_{k}^{(n)}(x_{1},x_{2},\ldots,x_{n})\sb.\label{generating}
\end{equation}
A general class of solution has been identified by Koubek and Mussardo
\cite{KM} and consists of the following expression 
\begin{equation}
Q_{n}(k)=\vert\vert M_{ij}(k)\vert\vert\sm,
\end{equation}
where $M_{ij}(k)$ is an $(n-1)\times(n-1)$ matrix with entries 
\begin{equation}
M_{ij}(k)=\s_{2i-j}\,[i-j+k]\sm.\label{element}
\end{equation}
These polynomials, which we call \emph{elementary solutions}, depend
on an arbitrary integer $k$ and satisfy 
\begin{equation}
Q_{n}(k)\,=\,(-1)^{n+1}Q_{n}(-k)\sm.\label{pr1}
\end{equation}
One can prove \cite{KM} that the whole set of FF of the elementary
field $\phi(x)$ are given by $Q_{n}(0)$ while those of the trace
of the energy-momentum tensor $\Theta(x)$ are given by the even polynomials
$Q_{2n}(1)$. One can also prove that these solutions of the FF equations
correspond to the matrix elements of the exponential operators $e^{kg\phi}$.

\section{Finite volume formalism \label{finvolFF}}

A formalism that gives the exact quantum form factors to all orders
in $L^{-1}$ was introduced in \cite{Pozsgay:2007kn,Pozsgay:2007gx}.
The finite volume multi-particle states can be denoted 
\begin{equation}
\vert\{I_{1},\dots,I_{n}\}\rangle_{L}
\end{equation}
where the $I_{k}$ are momentum quantum numbers, ordered as $I_{1}\geq\dots\geq I_{n}$
by convention. The corresponding energy levels are determined by the
Bethe-Yang equations 
\begin{equation}
\mathrm{e}^{imL\sinh\tilde{\beta}_{k}}\prod_{l\neq k}S(\tilde{\beta}_{k}-\tilde{\beta}_{l})=1
\end{equation}
We defining the two-particle phase shift $\delta(\beta)$ as 
\begin{equation}
S(\beta)=\mathrm{e}^{i\delta(\beta)}\label{deltadef}
\end{equation}
and its derivative will be denoted by 
\begin{equation}
\varphi(\beta)=\frac{d\delta(\beta)}{d\beta}\label{eq:phidef}
\end{equation}
Due to unitarity, $\delta$ is an odd and $\varphi$ is an even function.
We can write 
\begin{equation}
Q_{k}(\tilde{\beta}_{1},\dots,\tilde{\beta}_{n})=mL\sinh\tilde{\beta}_{k}+\sum_{l\neq k}\delta(\tilde{\beta}_{k}-\tilde{\beta}_{l})=2\pi I_{k}\quad,\quad k=1,\dots,n\label{eq:betheyang}
\end{equation}
where the quantum numbers $I_{k}$ take integer values. Eqns. \eqref{eq:betheyang}
must be solved with respect to the particle rapidities $\tilde{\beta}_{k}$,
where the energy (relative to the finite volume vacuum state) can
be computed as 
\begin{equation}
\sum_{k=1}^{n}m\cosh\tilde{\beta}_{k}
\end{equation}
up to corrections which decay exponentially with $L$. The density
of $n$-particle states in rapidity space can be calculated as 
\begin{equation}
\rho(\beta_{1},\dots,\beta_{n})_{L}=\det\mathcal{J}^{(n)}\qquad,\qquad\mathcal{J}_{kl}^{(n)}=\frac{\partial Q_{k}(\beta_{1},\dots,\beta_{n})}{\partial\beta_{l}}\quad,\quad k,l=1,\dots,n\label{eq:byjacobian}
\end{equation}
The finite volume behavior of local matrix elements can be given as
\cite{Pozsgay:2007kn} 
\begin{eqnarray}
\langle\{I_{1}',\dots,I_{m}'\}\vert\mathcal{O}(0,0)\vert\{I_{1},\dots,I_{n}\}\rangle_{L} & = & \frac{F_{m+n}^{\mathcal{O}}(\tilde{\beta}_{m}'+i\pi,\dots,\tilde{\beta}_{1}'+i\pi,\tilde{\beta}_{1},\dots,\tilde{\beta}_{n})}{\sqrt{\rho(\tilde{\beta}_{1},\dots,\tilde{\beta}_{n})\rho(\tilde{\beta}_{1}',\dots,\tilde{\beta}_{m}')}}\nonumber \\
 & + & O(\mathrm{e}^{-\mu L})\label{eq:genffrelation}
\end{eqnarray}
where $\tilde{\beta}_{k}$ ($\tilde{\beta}_{k}'$) are the solutions
of the Bethe-Yang equations (\ref{eq:betheyang}) corresponding to
the state with the specified quantum numbers $I_{1},\dots,I_{n}$
($I_{1}',\dots,I_{n}'$) at the given volume $L$. The above relation
is valid provided there are no disconnected terms i.e. the left and
the right states do not contain particles with the same rapidity,
i.e. the sets $\left\{ \tilde{\beta}_{1},\dots,\tilde{\beta}_{n}\right\} $
and $\left\{ \tilde{\beta}_{1}',\dots,\tilde{\beta}_{m}'\right\} $
are disjoint. In the course of work we do not need the relations containing
disconnected contributions; these were first obtained in \cite{Pozsgay:2007gx}.

In addition, the proper definition of a finite volume boundary state
was derived in \cite{Kormos:2010}. For a state of the form
\begin{equation}
|B\rangle=|0\rangle+\int\frac{d\beta}{2\pi}K(\beta)|-\beta,\beta\rangle+\frac{1}{2}\int\frac{d\beta_{1}d\beta_{2}}{2\pi}K(\beta_{1})K(\beta_{2})|-\beta_{1},\beta_{1},-\beta_{2},\beta_{2}\rangle+\dots
\end{equation}
it can be written as
\begin{align}
|B\rangle_{L} & =|0\rangle_{L}+\sum_{I}N_{2}(\tilde{\beta})_{L}K(\tilde{\beta})|\{-I,I\}\rangle_{L}\nonumber \\
 & +\sum_{I}N_{4}(\tilde{\beta}_{1},\tilde{\beta}_{2})_{L}K(\tilde{\beta}_{1},\tilde{\beta}_{2})|\{-I_{1},I_{1},-I_{2},I_{2}\}\rangle_{L}+\dots\label{eq:finvol_bstate}
\end{align}
where the rapidities $\{\tilde{\beta}_{1},\dots,\tilde{\beta}_{k}\}$
of the $2k$-particle term satisfy the appropriate finite volume quantization
conditions
\begin{align}
2\pi I_{l} & =\bar{Q}_{l}(\tilde{\beta}_{1},\dots,\tilde{\beta}_{k})\nonumber \\
 & =mL\sinh\tilde{\beta}_{l}+\sum_{k\neq l}\delta(\tilde{\beta}_{l}-\tilde{\beta}_{k})+\sum_{k}\delta(\tilde{\beta}_{l}+\tilde{\beta}_{k})\qquad l=1,\dots,k
\end{align}
The density of these states is given by
\begin{equation}
\bar{\rho}_{2k}(\tilde{\beta}_{1},\dots,\tilde{\beta}_{k})_{L}=\det\bar{\mathcal{J}}^{(k)}\qquad,\qquad\mathcal{\bar{J}}_{rs}^{(k)}=\frac{\partial\bar{Q}_{r}(\beta_{1},\dots,\beta_{k})}{\partial\beta_{s}}\quad,\quad r,s=1,\dots,k
\end{equation}
and the finite volume normalization coefficients are given by 
\begin{equation}
N_{2k}(\beta)_{L}=\frac{\sqrt{\rho_{2k}(-\beta_{1},\beta_{1},\dots-\beta_{k},\beta_{k})_{L}}}{\bar{\rho}_{2k}(\beta_{1},\dots,\beta_{k})_{L}}\label{eq:N2k}
\end{equation}

\section{Deriving the conditions satisfied by the boundary state}

The formalism described above is only valid up to finite size corrections
that decay exponentially with the volume. However, these have no effect
in the $L\rightarrow\infty$ limit and so we omit them below.

We demonstrate the method on the simple case of a Dirichlet boundary
state, which satisfies
\begin{equation}
\Phi(0)|D\rangle=0
\end{equation}
Substituting 
\begin{equation}
|D\rangle=|0\rangle+\int\frac{d\beta'}{2\pi}K(\beta')|-\beta',\beta'\rangle+\dots
\end{equation}
and taking the matrix element with a one-particle state, we obtain
\begin{equation}
\langle\beta|\Phi(0)|0\rangle+\int\frac{d\beta'}{2\pi}\langle\beta|\Phi(0)|-\beta',\beta'\rangle K(\beta')+\dots=0
\end{equation}
Using (\ref{eq:finvol_bstate}) this can be written as
\begin{equation}
\langle\{I\}|\Phi(0)|0\rangle_{L}+\sum_{I'}\langle\{I\}|\Phi(0)|\{-I',I'\}\rangle_{L}N_{2}(\beta')_{L}K(\beta')+\dots=0
\end{equation}
with the quantization conditions
\begin{eqnarray}
Q_{1}(\beta) & = & mL\sinh\beta=2\pi I\label{eq:oneptcond}\\
\bar{Q}_{2}(\beta') & = & mL\sinh\beta'+\delta(2\beta')=2\pi I'\label{eq:restrictedtwoptcond}
\end{eqnarray}
Neglecting the higher-order terms, we can then use eqns. (\ref{eq:genffrelation})
and (\ref{eq:finvol_bstate}) to write
\begin{equation}
\frac{F_{1}}{\sqrt{\rho_{1}(\beta)_{L}}}+\sum_{I'}\frac{F_{3}(\beta+i\pi,-\beta',\beta')}{\sqrt{\rho_{1}(\beta)_{L}}\sqrt{\rho_{2}(-\beta',\beta')_{L}}}N_{2}(\beta')_{L}K(\beta')+\dots=0
\end{equation}
where the state densities are
\begin{eqnarray}
\rho_{1}(\beta)_{L} & = & mL\cosh\beta\\
\rho_{2}(\beta_{1},\beta_{2})_{L} & = & mL\cosh\beta_{1}mL\cosh\beta_{2}+(mL\cosh\beta_{1}+mL\cosh\beta_{2})\varphi(\beta_{1}-\beta_{2})\nonumber 
\end{eqnarray}
Using (\ref{eq:N2k}), we can write 
\begin{equation}
F_{1}+\frac{1}{2}\sum_{\beta'}\frac{F_{3}(\beta+i\pi,-\beta',\beta')}{\bar{\rho}_{2}(\beta')_{L}}K(\beta')+\dots=0
\end{equation}
where we also extended the summation to negative values of rapidity.

Now one can trade the sum for a contour integral
\begin{equation}
\sum_{I'}\rightarrow\sum_{I'}\oint_{C_{I'}}\frac{d\beta'}{2\pi i}\frac{i\bar{\rho}_{2}(\beta')e^{i\bar{Q}_{2}(\beta')}}{e^{i\bar{Q}_{2}(\beta')}-1}
\end{equation}
where the contour $C_{I'}$ run around the solution of (\ref{eq:restrictedtwoptcond})
counter-clockwise, and so we obtain
\begin{equation}
F_{1}+\frac{1}{2}\sum_{I'}\oint_{C_{I'}}\frac{d\beta'}{2\pi}\frac{F_{3}(\beta+i\pi,-\beta',\beta')K(\beta')}{e^{i\bar{Q}_{2}(\beta')}-1}+\dots=0
\end{equation}
Open the contour into two lines running above and the below the real
axis to obtain
\begin{equation}
F_{1}+\frac{1}{2}\int_{-\infty}^{\infty}\frac{d\beta'}{2\pi}f(\beta'-i\epsilon)-\frac{1}{2}\int_{-\infty}^{\infty}\frac{d\beta'}{2\pi}f(\beta'+i\epsilon)-\frac{1}{2}\oint_{\beta}\frac{d\beta'}{2\pi}f(\beta')-\frac{1}{2}\oint_{-\beta}\frac{d\beta'}{2\pi}f(\beta')=0
\end{equation}
where the last two integrals are evaluated on contours surrounding
$\beta$ and $-\beta$, in counter-clockwise direction and
\begin{equation}
f(\beta')=\frac{1}{e^{i\bar{Q}_{2}(\beta')}-1}F_{3}(\beta+i\pi,\beta',-\beta')K(\beta')
\end{equation}
We can easily calculate 
\begin{equation}
\bar{Q}_{2}(\beta'\pm i\epsilon)=mL\left(\cos\epsilon\sinh\beta'\pm i\sin\epsilon\cosh\beta'\right)+O(L^{0})
\end{equation}
therefore
\begin{equation}
e^{i\bar{Q}_{2}(\beta')}\rightarrow\begin{cases}
0 & \,+\epsilon\\
\infty & \,-\epsilon
\end{cases}
\end{equation}
as $L\rightarrow\infty$ and
\begin{eqnarray}
\frac{1}{2}\int_{-\infty}^{\infty}\frac{d\beta'}{2\pi}f(\beta'-i\epsilon) & \rightarrow & 0\\
-\frac{1}{2}\int_{-\infty}^{\infty}\frac{d\beta'}{2\pi}f(\beta'+i\epsilon) & \rightarrow & \frac{1}{2}\int\frac{d\beta'}{2\pi}F_{3}(\beta+i\pi,-\beta'-i\epsilon,\beta'+i\epsilon)K(\beta'+i\epsilon)\nonumber 
\end{eqnarray}
Using the third form factor equation from (\ref{axioms}), we obtain
\begin{eqnarray}
F_{3}(\beta+i\pi,-\beta',+\beta') & = & (1-S(-2\beta'))\frac{iF_{1}}{\beta'-\beta}+\mbox{regular terms}\\
F_{3}(\beta+i\pi,-\beta',+\beta') & = & (1-S(-2\beta'))\frac{iF_{1}}{\beta'+\beta}+\mbox{regular terms}\nonumber 
\end{eqnarray}
and so the residue terms are
\begin{eqnarray}
 &  & -\frac{1}{2}\oint_{\beta}\frac{d\beta'}{2\pi}f(\beta')-\frac{1}{2}\oint_{-\beta}\frac{d\beta'}{2\pi}f(\beta')\\
 & = & -\frac{1}{4\pi}2\pi i\left(\mathop{\mathrm{Res}}_{\beta'=\beta}+\mathop{\mathrm{Res}}_{\beta'=-\beta}\right)f(\beta')\nonumber \\
 & = & -\frac{i}{2}\left\{ \frac{i(1-S(-2\beta))F_{1}K(\beta)}{e^{i\bar{Q}_{2}(\beta)}-1}+\frac{i(1-S(2\beta))F_{1}K(-\beta)}{e^{i\bar{Q}_{2}(-\beta)}-1}\right\} \nonumber 
\end{eqnarray}
From (\ref{eq:oneptcond},\ref{eq:restrictedtwoptcond}) one can evaluate
\begin{equation}
e^{\bar{Q}_{2}(\beta)}=S(2\beta)
\end{equation}
and so the pole contribution is
\begin{equation}
\frac{1}{2}F_{1}K(\beta)\left\{ 1+S(-2\beta)\right\} 
\end{equation}
where we also used the relations
\begin{eqnarray}
S(\beta)S(-\beta) & = & 1\\
S(2\beta)K(-\beta) & = & K(\beta)\nonumber 
\end{eqnarray}
Our final result for the integral equation to this order is
\begin{eqnarray}
0 & = & F_{1}+\frac{1}{2}F_{1}K(\beta)\left\{ 1+S(-2\beta)\right\} \\
 &  & +\frac{1}{2}\int_{-\infty}^{\infty}\frac{d\beta'}{2\pi}F_{3}(\beta+i\pi,-\beta'-i\epsilon,\beta'+i\epsilon)K(\beta'+i\epsilon)+\dots\nonumber 
\end{eqnarray}
This calculation can be extended to higher terms in the expansion
of the boundary state $|B\rangle$ using the multi-dimensional residue
method introduced in \cite{Pozsgay:2010cr}. One can also use any
other multi-particle test state in place of the one-particle state
applied above. Computation of these terms is lengthy and tedious,
but presents no further difficulties.

\section{Estimation of $\tau_0$ in a free bosonic model}

We consider a QQ of the mass in a relativistic free bosonic model. For this problem, the actual initial state is known exactly thus allowing the estimation of $\tau_0$. The initial state is
\eq[a1]{|\Psi\ra \sim \exp{\[-\int_0^\infty \frac{dp}{2\pi} \; \(\frac{E_{0p}-E_p}{E_{0p}+E_p}\) \;  A^\dg(-p) A^\dg(p) \]} |\Omega\ra}
where $E_p=\sqrt{p^2+m^2}$ is the energy dispersion relation (with $E_{0p}$ corresponding to mass $m_0$) and $A^\dg(p)$ is the post-quench bosonic creation operator. For $m_0 \to \infty$ this state tends to the Dirichlet state which therefore is $|D\ra \sim \exp{\(-\int_0^\infty \frac{dp}{2\pi} \; A^\dg(-p) A^\dg(p) \)} |0\ra $ and the approximation we want to use is
\eq[b1]{|\tilde\Psi\ra = e^{-H\tau_0}|D\ra \sim \exp{\(-\int_0^\infty \frac{dp}{2\pi} e^{-2E_p \tau_0} A^\dg(-p) A^\dg(p) \)} |\Omega\ra.}
In order to find an appropriate value for $\tau_0$ we should compare the expectation value of an observable $\mathcal{O}(\phi)$ in the exact and in the approximate state and require that they be equal. One choice of such an observable is the field fluctuations $\phi^2$. We therefore demand the condition $ \la \Psi|\phi^2|\Psi\ra = \la \tilde\Psi|\phi^2|\tilde\Psi\ra $ which reads 
\eq{ \int_0^\infty dp \, \frac{1}{E_{0p}} = \int_0^\infty dp \, \frac{\tanh (E_p \tau_0) }{E_p} }
leading for large $m_0$ (small $\tau_0$) to the following scaling relation
\eq[tau_0]{ \tau_0 = \frac{\pi}{2e^{\gamma}} m_0^{-1} \approx 0.881938 \; m_0^{-1} }
where $\gamma$ is the Euler-Mascheroni constant. Notice that the last expression is consistent with the conformal limit $m\to 0$ \cite{CC1,CC2} where we know that $\tau_0 \sim 1/m_0$, however the numerical prefactor is not the same.

\begin{figure}[htbp]
\centering
\includegraphics[width=.5 \columnwidth]{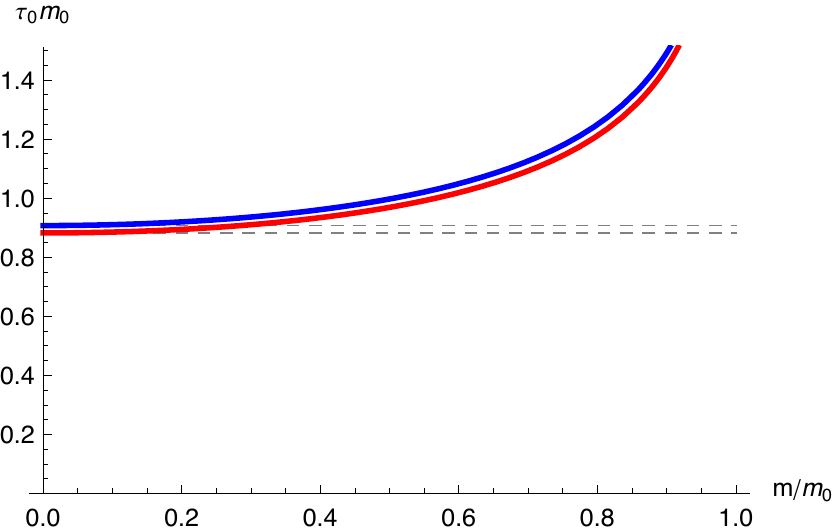}
\caption{ \label{fig} \small{Plot of $m_0 \tau_0$ as a function of $m/m_0$ as estimated based on the observables $\phi^2$ (red) and $H$ (blue curve). The dashed lines indicate the asymptotic values for $m_0\to\infty$.}}
\end{figure}

It is easy to see that the estimation of $\tau_0$ is \emph{not independent of the choice of observable} under consideration, not even in the large $m_0$ limit where $\tau_0$ should be uniquely defined if it is a good phenomenological parameter. Indeed let us repeat the estimation of $\tau_0$ using a different observable, in order to test if we obtain the same value. Since the states we consider are gaussian, all higher moments of the field fluctuations $\la\phi^{2n}\ra$ are simply powers of the variance $\la\phi^{2n}\ra = (2n-1)!! \la\phi^2\ra^n$ and therefore by fixing $\tau_0$ from the variance as above, we fix all higher moments too. Therefore using any of these higher moments to obtain $\tau_0$ would automatically result in the same estimate. Another observable of interest is the energy density. By requiring that $ \la \Psi|H|\Psi\ra = \la \tilde\Psi|H|\tilde\Psi\ra $ we have 
\eq{ \int_0^\infty dp \, \frac{E_{0p}^2+E_p^2}{2E_{0p}}= \int_0^\infty dp \, {E_p} \coth (2 E_p \tau_0) }
which finally gives
\eq[tau_0b]{\tau_0 = \frac{\pi}{2\sqrt{3}} m_0^{-1} \approx 0.906900 \; m_0^{-1} }
This estimate differs from (\ref{tau_0}) in the numerical factor by about 3\%. 
Fig.~\ref{fig} shows $\tau_0 m_0$ as a function of $m/m_0$ as estimated based on the field fluctuations $\phi^2$ and the energy density $H$ by numerical evaluation of the corresponding equations. The dependence of $\tau_0$ on the observable is rather expected and can be explained by that fact that a QQ creates excitations of arbitrarily high energy and therefore cannot be analysed by means of the low-energy physics only.

Perhaps a more serious flaw comes up when we compare the exact with the approximate large time asymptotic value of an observable, as derived based on the value $\tau_0$ estimated from this same observable. For the field fluctuations our approximation turns out to give $ \lim_{t\to\infty} \la\tilde\Psi|\phi^2(t) |\tilde\Psi\ra_{(R)} = {\pi}/{8m\tau_0} $ (the subscript $(R)$ stands for ``renormalised'', i.e. after mass renormalisation has been applied) whereas the exact value is $ \lim_{t\to\infty} \la\Psi|\phi^2(t) |\Psi\ra_{(R)} = {\pi m_0}/{8m} $, always in the limit of large $m_0$, small $\tau_0$. The two results would be equal to each other if $\tau_0$ was simply equal to $m_0^{-1}$ instead of (\ref{tau_0}). Therefore our estimate (\ref{tau_0}) based on the comparison of the initial field fluctuations \emph{does not reproduce the exact results consistently} as far as numerical factors are concerned.

As suggested in the introduction, in order to overcome these failures of the $\tau_0$ regularisation, one may assume a momentum dependent $\tau_0$ so that (\ref{a1}) reproduces exactly (\ref{b1}), i.e. by defining
\eq{ \tau_0(p) \equiv -\frac{1}{2 E_p} \log{\(\frac{E_{0p}-E_p}{E_{0p}+E_p}\)}}
Notice that for large $m_0$, the above equation gives $\tau_0 \sim 1/E_{0p} \sim 1/m_0$ as expected.

\section{Numerical solution of the integral equations (19) and (21) given in the text}

In order to numerically solve the integral equations (19) and (21) we employ the iterative method. Truncated after the first 5 terms, the equations are of the form 
\eq{ K(\theta) = K_f(\theta) + \int d\theta' \, G(\theta,\theta') K(\theta') + \int d\theta'_1 d\theta'_2 \, G(\theta,\theta'_1,\theta'_2) K(\theta'_1) K(\theta'_2) }
which are non-linear integral equations. If we further omit the 4th and 5th terms, the equations reduce to linear inhomogeneous Fredholm integral equations of the 2nd kind. In the iterative method we start by using a guess or approximation of the solution $K_{(0)}(\theta)$ (for example $K_{(0)}(\theta) = K_f(\theta)$ or $K_{(0)}(\theta) = K_D(\theta)$ in the case of (19)) and calculate the first correction 
\eq{ K_{(1)}(\theta) = K_f(\theta) + \int d\theta' \, G(\theta,\theta') K_{(0)}(\theta') + \int d\theta'_1 d\theta'_2 \, G(\theta,\theta'_1,\theta'_2) K_{(0)}(\theta'_1) K_{(0)}(\theta'_2)}
We proceed similarly to calculate the second correction $K_{(2)}(\theta)$ from $K_{(1)}(\theta)$ and continue iterating this step until the new correction is sufficiently close to the previous one. The convergence criterion can be chosen to be that the difference of two successive corrections is everywhere smaller than some chosen tolerance. To ensure or make convergence faster we can mix every new correction with the previous one using suitable weights.

A complication arises in the above general iterative procedure by the fact that the integration should be performed not along the real $\theta$-axis but along a parallel line shifted by $i\epsilon$. This means that an iteration step does not give $K_{(n)}$ along the same line where $K_{(n-1)}$ is known but a shifted one and therefore the iteration breaks. There are several ways to solve this problem. One way is to set $\epsilon$ to a very small value so that we can assume that the obtained function $K_{(n)}(\theta)$ is approximately equal to $K_{(n)}(\theta+i\epsilon)$ which is needed in the next iteration step. The value of $\epsilon$ has to be sufficiently small so that the cumulative error after all iterations needed in order to reach convergence is small. An alternative way to apply this approximation is to consider the positions of the kinematical poles shifted according to the $\epsilon$-prescription and bring the integration contour along the real axis, so that the iterative method can be applied without problem. In this approximation the integration kernels can be simplified using the analytical properties of the form factors and can be expressed in terms of only one numerically evaluated function: the function $F_{\text{min}}(\beta+i\pi)$ for real $\beta$ which is smooth and easily calculated numerically. Another approach is to consider two copies of the integral equations for two opposite values of $\epsilon$ so that at each iteration we calculate $K_{(n)}(\theta-i\epsilon)$ from $K_{(n-1)}(\theta+i\epsilon)$ and $K_{(n)}(\theta+i\epsilon)$ from $K_{(n-1)}(\theta-i\epsilon)$, therefore obtaining all the information needed for the next iteration step. The advantage of this approach is that the value of $\epsilon$ does not have to be very small; simply smaller than the distance of the closest singularity of the integrands to the real axis. The solution can be found along the real axis after the last step. 

Another practical point is that, since the 5th term of the equation requires more computing time due to the double rapidity integral that it contains but is smaller in size than the previous terms, one may choose a less dense discretisation of rapidity to reduce the computing time. In addition one may omit this term from the equation at first, solve it and then improve the solution by including the previously omitted term and solving again. This two-stage iterative method obviously requires less overall computing time.

In Fig.~2 and 3, we have used the first approach for the numerical implementation of the $\epsilon$-prescription, i.e. we used a very small value $\epsilon=0.0001$ and ignored the line shifts. As a starting guess we used $K_D(\theta)$ for equation (19) and our ansatz (22) for equation (21). We iterate the equation first omitting the 4th and 5th terms which we include later. Convergence of the iterative procedure is achieved with a tolerance of the order of $0.05$ after only one step almost everywhere; except the neighbourhood of the point $\theta=0$ where pathological behaviour should be expected due to the fact that the truncation of the form factor series becomes worse at this point.

\end{document}